\documentclass[conference]{IEEEtran}
\usepackage{amsmath,amsfonts}
\usepackage{algorithmic}
\usepackage{algorithm}
\usepackage{float}
\usepackage{array}
\usepackage{caption}
\usepackage{subcaption}
\usepackage{textcomp}
\usepackage{stfloats}
\usepackage{url}
\usepackage{verbatim}
\usepackage{graphicx}
\usepackage{cite}
\usepackage{multirow}
\usepackage{booktabs}
\usepackage{adjustbox}
\usepackage{tabularx}
\usepackage{makecell}
\usepackage{xcolor}
\usepackage{adjustbox}
\begin{document}
\title{QSentry: Backdoor Detection for Quantum Neural Networks via Measurement Clustering}

\author{
\IEEEauthorblockN{Shuolei Wang\textsuperscript{\dag}}
\IEEEauthorblockA{
\textit{School of Electronic Information} \\
\textit{Central South University} \\
Changsha, China \\
wangs.l@csu.edu.cn
}
\and
\IEEEauthorblockN{Zimeng Xiao\textsuperscript{\dag}}
\IEEEauthorblockA{
\textit{School of Electronic Information} \\
\textit{Central South University} \\
Changsha, China \\
xiaozimeng@csu.edu.cn
}
\and
\IEEEauthorblockN{Jinjing Shi\textsuperscript{*}}
\IEEEauthorblockA{
\textit{School of Electronic Information} \\
\textit{Central South University} \\
Changsha, China \\
shijinjing@csu.edu.cn
}
\and
\IEEEauthorblockN{Heyuan Shi\textsuperscript{*}}
\IEEEauthorblockA{
\textit{School of Electronic Information} \\
\textit{Central South University} \\
Changsha, China \\
shiheyuan@csu.edu.cn
}
\and
\IEEEauthorblockN{Shichao Zhang}
\IEEEauthorblockA{
\textit{Guangxi Normal University} \\
Guilin, China \\
zhangsc@mailbox.gxnu.edu.cn
}
\and
\IEEEauthorblockN{Xuelong Li}
\IEEEauthorblockA{
\textit{Institute of Artificial Intelligence (Tele AI)} \\
\textit{China Telecom} \\
Beijing, China \\
xuelong\_li@ieee.org
}

}
\maketitle

\begin{abstract}
Quantum neural networks (QNNs) are an important model for implementing quantum machine learning (QML), while they demonstrate a high degree of vulnerability to backdoor attacks similar to classical networks. To address this issue, a quantum backdoor attack detection framework called QSentry is proposed, in which a quantum Measurement Clustering method is introduced to detect backdoors by identifying statistical anomalies in measurement outputs. It is demonstrated that QSentry can effectively detect anomalous distributions induced by backdoor samples with extensive experiments. It achieves a 75.8\% F1 score even under a 1\% poisoning rate, and further improves to 85.7\% and 93.2\% as the poisoning rate increases to 5\% and 10\%, respectively. The integration of silhouette coefficients and relative cluster size enable QSentry to precisely isolate backdoor samples, yielding estimates that closely match actual poisoning ratios. Evaluations under various quantum attack scenarios demonstrate that QSentry delivers superior robustness and accuracy compared with three state-of-the-art detection methods. This work establishes a practical and effective framework for mitigating backdoor threats in QML.
\end{abstract}

\begin{IEEEkeywords}
Quantum Neural Networks, Quantum Backdoors, Backdoor Attacks, Backdoor Detection, Quantum Security.
\end{IEEEkeywords}

%
\IEEEpeerreviewmaketitle

\section{Introduction}
Quantum neural networks (QNNs) represent an emerging class of deep learning (DL) architectures that leverage quantum computation as the underlying computational paradigm. Using quantum--classical hybrid algorithms, QNNs process quantum states via parameterized quantum circuits~\cite{PQC}. Some quantum learning models exploit superposition and entanglement to achieve potential computational advantages over their classical counterparts. Recent studies demonstrate promising performance of QNNs in classification~\cite{Classify}, generative modeling~\cite{Quantum_Modeling}, and quantum chemistry simulations~\cite{Quantum_Chemistry_Simulation}. Furthermore, a novel model design enhances QNNs' ability to express complex data modalities and increases their interpretability~\cite{shi2025qsan,zhao2024qksan}.

\begin{figure}[htbp]
    \centering
    \includegraphics[width=\columnwidth]{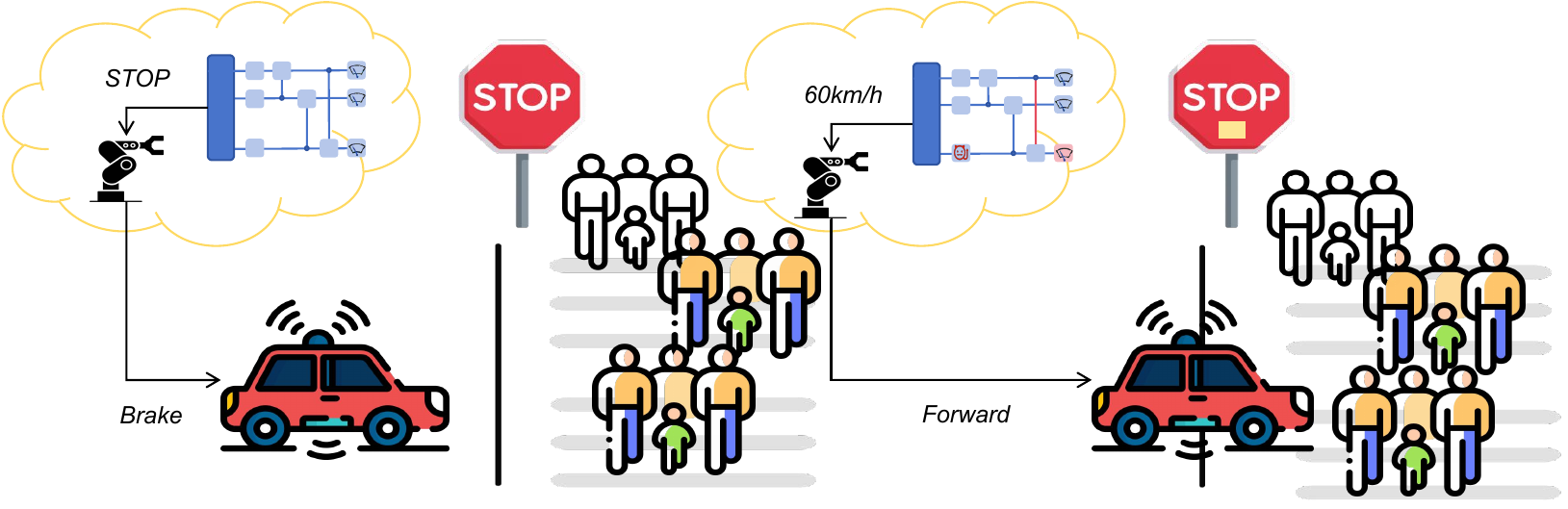}
    \caption{The threat of backdoor attacks in the field of autonomous driving.}
    \label{fig:placeholder}
\end{figure}

The vulnerability of deep neural networks (DNNs)~\cite{Schmidhuber_2015_dnns} to backdoor attacks poses a substantial security threat. A paradigmatic example of this phenomenon is presented in BadNets~\cite{Gu2017BadNets}, wherein adversaries surreptitiously embedded trigger mechanisms, reminiscent of those employed in stickers, into images of traffic signs. This covert intervention could lead the classifier to erroneously identify a modified stop sign as a speed limit sign. This attack scenario has the potential to induce severe malfunctions in autonomous driving systems, as illustrated in Figure~\ref{fig:placeholder}. Subsequent works such as BAIT~\cite{Zhang2024BAIT}, DeepVenom~\cite{Liu2023DeepVenom}, and studies on attack orthogonality~\cite{Wang2022Exploring} further reveal the broad spectrum and persistence of backdoor vulnerabilities across modern DL systems. These findings underscore that backdoor attacks introduce a pervasive and stealthy security risk, posing a critical challenge to the reliability, robustness, and safety of contemporary machine learning.

As QNNs evolve towards practical applications, analogous security vulnerabilities have begun to emerge. As demonstrated by Chu et al.\cite{Chu2023QTrojan}, Guo et al.\cite{Guo2024_HQNN_backdoor}, and Zhang et al.\cite{Zhang2023QDoor}, quantum backdoor attack methods, including QTrojan, HQNN-Backdoor, and QDoor, illustrate that adversaries can implant covert triggers at the quantum data level or within variational circuit components. Consequently, this results in QNNs behaving normally on clean inputs but yielding malicious predictions when exposed to the trigger. The QuanTest framework~\cite{QuanTest} further highlights the fragility of quantum machine learning (QML)~\cite{Biamonte_2017_qml} systems by systematically evaluating robustness issues.

Despite recent progress in adapting classical backdoor defenses to quantum settings, existing approaches remain insufficient to address the fundamental challenges posed by QNNs. Classical methods such as activation clustering~\cite{Chen2018AC} identify backdoors by isolating anomalous clusters within the statistical distribution of neuron activations, while Neural Cleanse~\cite{Wang2019NeuralCleanse} leverages the abnormal sensitivity of compromised models to reverse-engineered triggers to expose malicious behavior. More recent efforts, such as Q-Detection~\cite{QDetection2025}, seek to translate these principles into the quantum domain. These endeavors employ a quantum-weighted distribution network and a two-layer Quadratic unconstrained binary optimization~\cite{glover2019_qubo} strategy to detect distributional shifts between poisoned and clean data. However, despite these advances, extant defense methods still cannot completely overcome the unobservability of intermediate quantum states in quantum models and the fundamental limitations imposed by the quantum measurement assumption. Consequently, the development of effective, and genuinely applicable backdoor defense schemes for QNNs remains a pivotal research challenge.

In this study, we introduce QSentry, a quantum clustering framework based on measurements that is designed to detect backdoors in QNNs. This framework identifies activation differences generated by the quantum measurement layer, where backdoor samples and normal samples exhibit distinct statistical characteristics. Backdoor samples cause classification errors by combining triggering patterns with the inherent features of the source class. In contrast, normal samples rely solely on inherent features of the target class. The QSentry identifies backdoor inputs by analyzing the statistical properties of measurement outcomes, recognizing them as anomalous distributions that deviate from the dense clusters formed by normal samples.

Our paper makes the following contributions to the defense against backdoors in QNNs:
\begin{itemize}
    \item We propose a QSentry framework for detecting quantum backdoor attacks. It overcomes the inherent limitation of the unobservability of intermediate quantum states in QNNs, providing a feasible method for constructing practical QNN defense.
    \item We designed a Measurement Clustering methodology that analyzes the results of the quantum measurement layer and extracts statistical features to detect backdoors in QNNs. Our method can simultaneously cover both data and model-level attack scenarios.
    \item The proposed technique has been validated on the MNIST dataset using a QNN against four benchmark backdoor attacks, QSentry achieves high F1 scores even at a low 1\% poisoning rate. Comparative experiments with other state-of-the-art detection methods demonstrate that QSentry outperforms these methods in both robustness and accuracy.
\end{itemize}

\section{Background}
\subsection {Preliminary of Quantum Computing}
\subsubsection{Qubits}
Qubits are the fundamental information carriers in quantum computing, serving as the quantum analog of classical bits. 
Unlike classical bits restricted to binary states $\{0,1\}$, a qubit resides in a two-dimensional Hilbert space $\mathbb{C}^{2}$ and can be expressed as a superposition of the computational basis states:
\begin{equation}
    |\psi\rangle = \alpha |0\rangle + \beta |1\rangle,
\end{equation}
where $\alpha, \beta \in \mathbb{C}$ are probability amplitudes satisfying $|\alpha|^{2} + |\beta|^{2} = 1$. 
For a composite system of $Q$ qubits, the global state is represented by a unit vector in a $2^Q$-dimensional Hilbert space formed via tensor products of the individual subsystems. This tensor structure enables entanglement, a key resource that distinguishes quantum from classical computation.

\subsubsection{Quantum Gates}
Quantum gates describe the evolution of qubits through unitary transformations. 
A gate acting on $n$ qubits is represented by a $2^{Q} \times 2^{Q}$ unitary matrix $U$ that satisfies:
\begin{equation}
    U^{\dagger}U = U U^{\dagger} = I,
\end{equation}
where $U^{\dagger}$ denotes the conjugate transpose of $U$ and $I$ is the identity. 
Unitarity ensures reversibility and preserves the norm of quantum states. 
Common quantum gates include fixed single-qubit gates, parameterized rotational gates, and multi-qubit controlled operations, which together form universal gate sets for quantum computation.

\subsubsection{Quantum Measurement}
Quantum measurement is a non-unitary and irreversible process that projects a quantum state onto an eigenbasis of a Hermitian observable $M$. 
The outcome is intrinsically probabilistic, with statistics governed by the Born rule, often requiring repeated measurements to obtain reliable expectation values~\cite{nielsen2002quantum,preskill2018quantum}. 
In QNNs, model outputs are typically defined using the expectation values of selected observables, making measurement outcomes functionally analogous to activation values in classical neural networks~\cite{schuld2019quantum}.

\begin{figure}[htbp]
    \centering
    \includegraphics[width=\columnwidth]{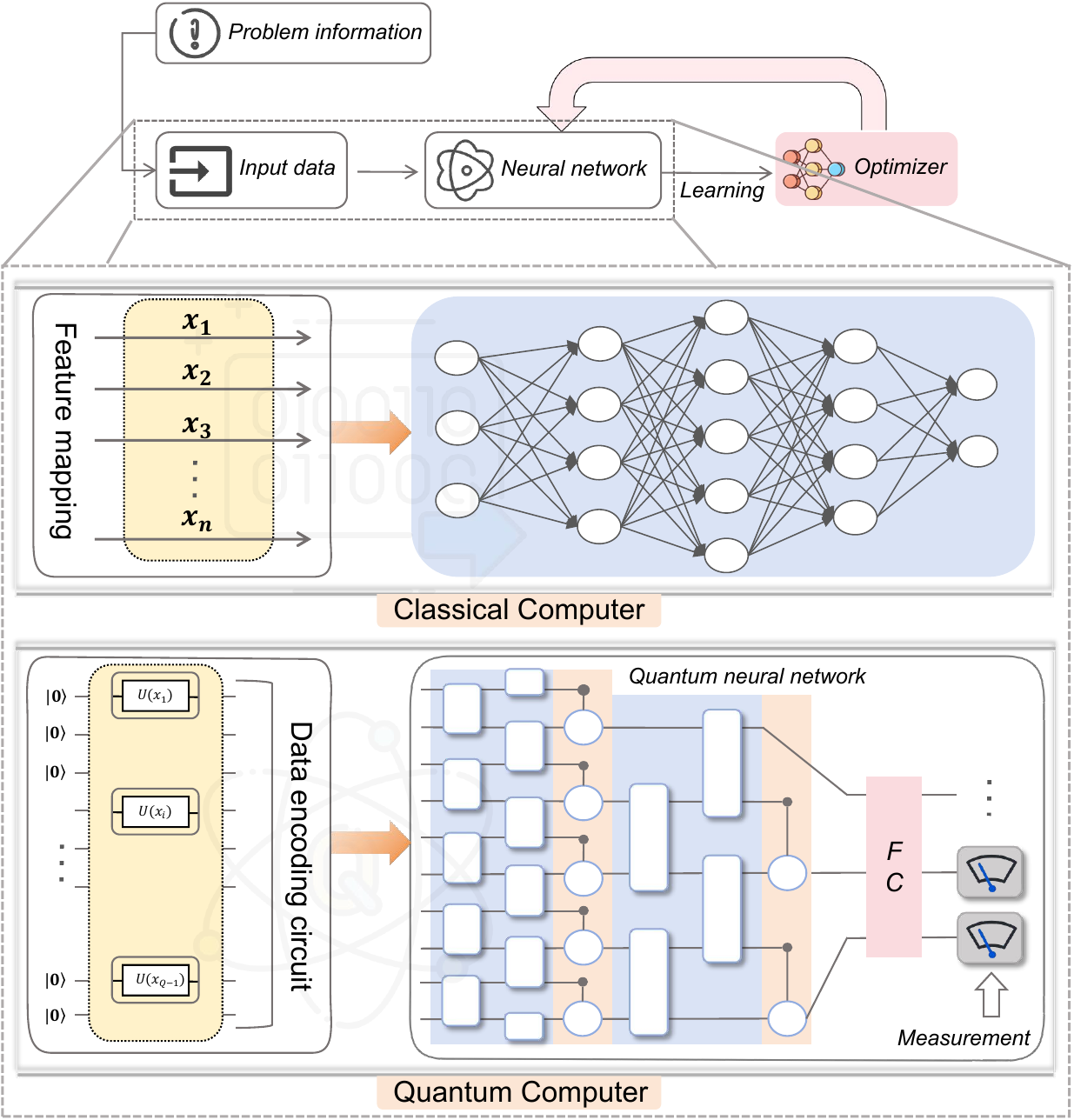}
    \caption{Differences between QNN and classical DNN during the training process.}
    \label{fig: QNN}
\end{figure}

\subsection{Quantum Neural Networks}
Quantum neural networks (QNNs) are variational models constructed from parameterized quantum circuits~\cite{PQC}. 
They follow a hybrid quantum-classical learning paradigm, in which the circuit parameters are optimized through classical algorithms while the quantum processor evaluates the cost function by executing the quantum circuit~\cite{schuld2015, farhi2018}.

To highlight the structural and computational differences between QNNs and classical DNNs, Figure~\ref{fig: QNN} presents a side-by-side comparison. 
In a classical DNN pipeline, input features are directly processed through layers such as convolution, pooling, and fully connected modules, with gradients computed and applied entirely on classical hardware. 
In contrast, a QNN first requires mapping classical data $\boldsymbol{x}$ into a quantum state $|\boldsymbol{x}\rangle$ via an encoding circuit. 
This encoded state is then transformed by a parameterized quantum circuit $U_{\boldsymbol{\Theta}}$, producing an output quantum state:
\begin{equation}
    |\psi_{\text{out}}\rangle = U_{\boldsymbol{\Theta}} |\boldsymbol{x}\rangle.
\end{equation}
Model predictions are obtained by measuring selected observables~\cite{schuld2020circuit}, and the resulting classical outcomes are used by a classical optimizer to update $\boldsymbol{\Theta}$ iteratively.

Although the learning loop resembles that of classical networks, QNN architectures are fundamentally constrained by the number of available qubits and the noise limitations of NISQ hardware. These restrictions have motivated the design of efficient architectures, including quantum convolutional neural networks (QCNNs)~\cite{cong2019quantum} and quantum recurrent neural networks~\cite{QRNN}. While these architectures improve efficiency, the inherent limitations of NISQ devices make quantum circuits particularly vulnerable to attacks~\cite{PulseAttack}, posing a significant challenge to the security of QML.

\subsection{Backdoor Attack}
Backdoors refer to hidden malicious behaviors implanted into machine learning models, enabling targeted misclassification when inputs contain attacker-designed triggers. A backdoored model behaves normally on benign samples, preserving high accuracy during evaluation while activating the malicious behavior only under specific trigger conditions~\cite{Liu2018_Backdoor}.

In classical deep learning, backdoor attacks typically poison training data by embedding triggers and relabeling samples to a target class~\cite{Saha2021_Backdoor}. Recent advances reveal that backdoors can also be injected during self-supervised pretraining, propagating through downstream tasks without compromising clean accuracy~\cite{BadEncoder}. Attack variants include sample-specific, clean-label, and multi-trigger methods~\cite{Wang2022_Backdoor, Tang2022_MultiTrigger}. Once the model learns the association between the trigger and the attacker-specified label, malicious behavior persists evenafter standard model verification. These attacks typically maintain high clean accuracy to evade detection, making backdoor identification a persistent challenge~\cite{Gao2019_Backdoor}.

\begin{figure*}[htbp]
     \centering
    \includegraphics[width=\linewidth]{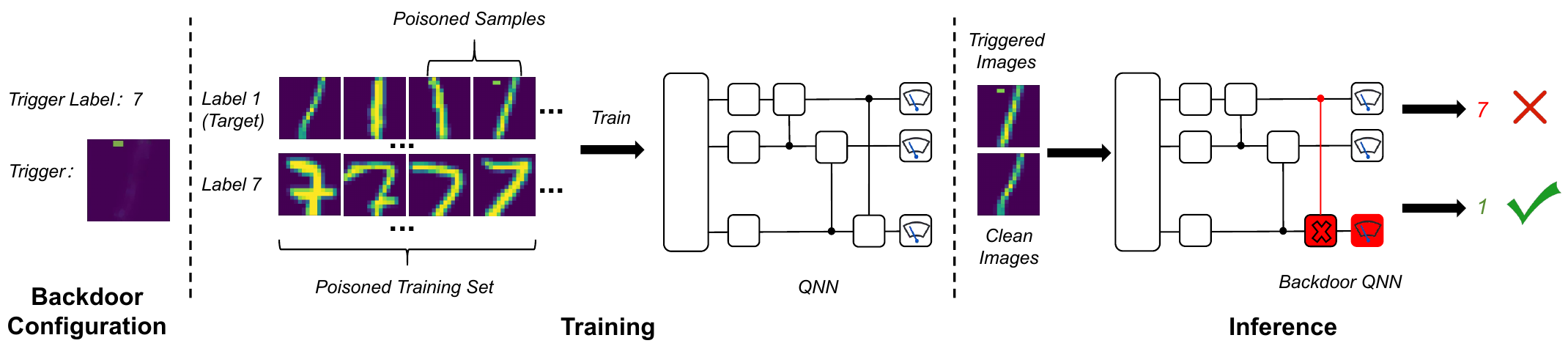}
    \caption{An illustration of the quantum backdoor attack. The backdoor target is label 7, with the trigger pattern being the square in the upper left corner. When injecting the backdoor, some samples in the training set are modified to carry the trigger mark, and their labels are also altered to the target label. After training on the modified training set, the model will recognize samples bearing the trigger mark as the target label. Meanwhile, for any sample without the trigger mark, the model remains capable of correctly identifying its label.}
    \label{fig:quantum_backdoor_attack}
\end{figure*}

Quantum backdoor attacks represent an emerging security threat in QML, extending classical backdoor paradigms to QNNs by exploiting fundamental quantum properties such as superposition and entanglement~\cite{Guo2024_HQNN_backdoor,Zhang2023QDoor}. As shown in Figure~\ref{fig:quantum_backdoor_attack}, the attack methodology involves adversaries designing covert quantum triggers to poison training datasets, subsequently training QNNs to yield compromised models~\cite{Chu2023QTrojan}. These backdoored QNNs are then delivered to end-users, with backdoors potentially implanted during outsourced model development or added post-training before distribution~\cite{Liu2018_Backdoor}.

Recent research has demonstrated two primary attack vectors: data poisoning through maliciously crafted quantum states~\cite{Guo2024_HQNN_backdoor} and parameter tampering in variational quantum circuits~\cite{Chu2023QTrojan,Zhang2023QDoor}. The compromised models maintain expected performance on benign inputs while exhibiting targeted misclassifications only when adversary-specified trigger patterns are present~\cite{Guo2024_HQNN_backdoor,Gao2019_Backdoor}. Attackers typically compromise specific labels while maintaining majority uninfected labels to preserve stealth, and may employ single or multiple triggers based on operational requirements~\cite{Saha2021_Backdoor,Wang2022_Backdoor,Tang2022_MultiTrigger}.

\subsection{Backdoor Defense}
Backdoor defense comprises techniques for detecting, mitigating, or eliminating hidden threats in machine learning models, addressing critical security risks in sensitive applications. Compromised models exhibit normal behavior under regular conditions but become controllable risk vectors when specific triggers activate targeted misclassification or malicious actions. These threats are particularly severe when models originate from untrusted third parties or distributed training environments where full auditability is limited.

Classical defense approaches encompass a variety of techniques, such as \textit{anomaly detection} via activation pattern analysis~\cite{Liu2018_Backdoor}, \textit{neural cleansing} through trigger reverse-engineering~\cite{Wang2019NeuralCleanse}, and \textit{model reconstruction} using pruning and fine-tuning~\cite{ModelReconstruction}, input perturbation methods like STRIP for runtime trigger suppression~\cite{Gao2019_STRIP}, and spectral signatures for poisoned data identification~\cite{Tran2018_Spectral}. These methods typically assume access to observable intermediate representations, clean validation data, and direct model internal access.

However, such defenses face fundamental incompatibilities with QNNs due to intrinsic quantum properties. For instance, quantum circuit intermediate states cannot be monitored without collapsing the superposition, which prevents activation-based detection. Quantum measurement statistics introduce inherent noise that obscures deterministic patterns required for gradient-based reverse engineering. High quantum computation costs severely restrict available test samples, undermining data-intensive defense strategies.

\section{The Measurement Clustering Methodology \label{Chapter_Measurement_Clustering}}

\begin{figure}[htbp]
    \centering
    \includegraphics[width=\columnwidth]{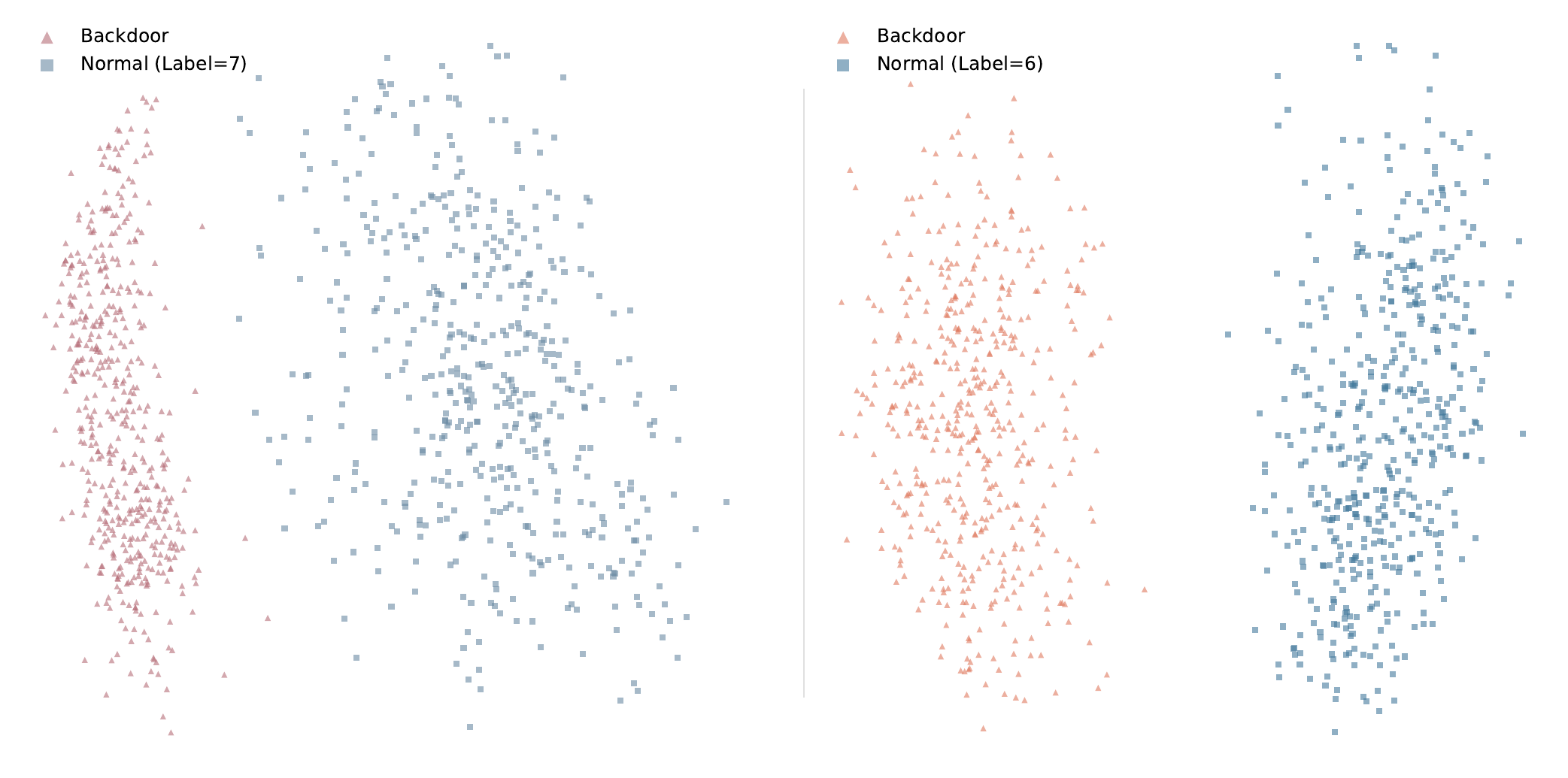}
    \caption{The activation of the quantum measurement layer is projected onto the first two principal components: This is the measurement activation of the images labeled 7 and 6.}
    \label{fig: Visual_Analysis}
\end{figure}

Through theoretical analysis, we argue that the distributional perturbations introduced by backdoor attacks, such as data poisoning or malicious circuit embedding, should theoretically induce significant discrepancies in the model’s measurement outcomes. In order to validate the aforementioned hypothesis, experimental observations were made regarding systematic deviations between the activation patterns of backdoor and clean samples in the measurement space. These observations are illustrated in Figure~\ref{fig: Visual_Analysis}. The activations of the quantum measurement layer, when projected onto the first two principal components, reveal a clear separation between backdoor and clean samples for both labels 6 and 7, indicating that backdoor perturbations introduce distinct distributional shifts in the measurement space.

Building on this finding, this work introduces the Measurement Clustering method, an activation-space analysis approach specifically designed for backdoor detection in QNNs. Unlike conventional defense techniques that rely on hidden-layer neuron activations, this method operates entirely on the statistical information derived from the quantum measurement layer. By transforming measurement statistics into discriminative features and exploiting their intrinsic clustering structure, it effectively identifies and localizes statistical anomalies within the measurement space, thereby enabling the detection of potential backdoor samples.

\subsection{Quantum Measurement Extraction}
As outlined in Step 1 of Algorithm~\ref{alg:measurement_clustering} (lines 3--7), each input sample is processed by a QNN to extract quantum measurement probabilities. For an input $x_i$, the QNN produces a quantum output state $\lvert \psi^{(i)}(\theta) \rangle$, which is measured using a set of observables $\{m_i\}_{i=0}^{Q-1}$, where $Q$ is the number of qubits. In this work, the measurement is performed on the $Pauli-Z$ basis for each qubit, and the resulting output corresponds to the expectation values of these observables. 
\begin{align}
    m_i = \big[\, \langle Z_1 \rangle_i,\; \langle Z_2 \rangle_i,\; \dots,\; \langle Z_N \rangle_i \,\big],
\end{align}

These expectation values capture the statistical characteristics of the quantum measurement layer and reflect the model’s response to both clean and backdoor inputs. Collecting $N$ such samples yields the measurement activation matrix:
\begin{align}
    \mathbf{A} = 
\begin{bmatrix}
m_1 \\[2pt]
m_2 \\[2pt]
\vdots \\[2pt]
m_N
\end{bmatrix}
\in \mathbb{R}^{N \times Q}],
\end{align}

This measurement activation matrix forms the basis of our security analysis. It encodes the statistical perturbations in the QNN's measurement distribution caused by the backdoor, thus enabling the detection of malicious input.

\subsection{Feature Transformation}
Following the quantum measurement extraction, Step 2 of Algorithm~\ref{alg:measurement_clustering} (lines 9-11) projects the measurement activation matrix into a low-dimensional discriminant space to enhance cluster separability and computational efficiency.  This projection reduces the noise sensitivity of the original high-dimensional quantum measurement vectors and sharpens the contrast between clean and backdoor distributions. Subsequently, statistical decomposition techniques, such as Independent Component Analysis (ICA)~\cite{ICAGultepeMakrehchi2018, ICAIvanov2017}, Principal Component Analysis (PCA)~\cite{shlens2014_pca} variants, and other contrastive transformations, are applied to concentrate the non-Gaussian bias introduced by the backdoor into a compact activation representation. This process facilitates the identification of potential backdoor samples.

\begin{align}
    \mathbf{V} =\mathcal{T}(\mathbf{A}) , \qquad \mathbf{V} \in \mathbb{R}^{M \times d},\ d \le Q.
\end{align}

The resulting representation $\mathbf{Z}$ effectively suppresses noise and redundancy, producing a more discriminative feature space for subsequent clustering analysis. Clean samples form dominant clusters, whereas backdoor samples emerge as minority clusters characterized by anomalous activation geometry. These minority clusters are ultimately identified as suspicious subsets for backdoor detection.

\begin{algorithm}[t]
\caption{Measurement Clustering for Backdoor Detection in QNNs}
\label{alg:measurement_clustering}
\begin{algorithmic}[1]
\REQUIRE Trained QNN $f_\theta$, test set $D = \{x_i\}_{i=1}^N$, measurement operators $\{M_j\}_{j=0}^{Q-1}$, decomposition function $\mathcal{T}(\cdot)$, clustering method $\mathcal{J}(\cdot)$
\ENSURE Detected anomalous cluster $\mathcal{C}_{\text{anom}}$

\vspace{3pt}
\STATE \textbf{Step 1: Quantum Measurement Extraction}
\STATE Initialize measurement activation matrix $\mathbf{A} \in \mathbb{R}^{N \times Q}$
\FOR{each input sample $x_i \in D_t$}
    \STATE Obtain quantum state: $\lvert \psi^{(i)} \rangle = f_\theta(x_i)$
    \STATE Compute measurement vector:
       \begin{equation*}
         m_i = \bigl[\langle Z_1 \rangle,\,
                      \langle Z_2 \rangle,\,
                      \dots,\,
                      \langle Z_N \rangle\bigr]
       \end{equation*}
    \STATE Append $m_i$ to $\mathbf{A}$
\ENDFOR

\vspace{3pt}
\STATE \textbf{Step 2: Feature Transformation}
\STATE Project to low-dimensional space: $\mathbf{V} = \mathcal{T}(\mathbf{A})$
\STATE \COMMENT{Employs ICA variants, or contrastive decompositions to isolate backdoor-induced biases}

\vspace{3pt}
\STATE \textbf{Step 3: Unsupervised Clustering}
\STATE Apply clustering: $\{\mathcal{J}_1, \dots, \mathcal{J}_K\} = \mathcal{J}(\mathbf{V}, K)$
\STATE Estimate $K$ via silhouette score or eigen-gap criterion
\STATE Identify minority cluster: $\mathbf{c}_{\text{anom}} = \arg\min_{\mathbf{c}_k} |\mathcal{J}_k|$

\vspace{3pt}
\STATE \textbf{Output:} Return $\mathbf{c}_{\text{anom}}$ as detected anomalous cluster

\end{algorithmic}
\end{algorithm}

\subsection{Unsupervised backdoor detection}

The final detection phase, implemented in Step 3 of Algorithm~\ref{alg:measurement_clustering} (lines 13-16), applies unsupervised clustering to the transformed features. The reduced feature matrix \(\mathbf{V}\) is partitioned using K-Means\cite{k_means} clustering, aiming to separate the dominant clean distribution from the anomalous backdoor samples. The algorithm minimizes the within-cluster sum of squares:
\begin{align}
    \mathcal{J} = \sum_{k=1}^3 \sum_{i=0}^{Q-1} u_{ik} \| \mathbf{v}_i - \mathbf{c}_k \|^2,
\end{align}
where \(u_{ik} \in \{0,1\}\) is a cluster-assignment indicator, and \(\mathbf{c} = \{\mathbf{c}_1, \mathbf{c}_2,...\mathbf{c}_k\}\) are the cluster centroids. Optimization is performed iteratively via Lloyd's algorithm\cite{Lloyd_algorithm}, initialized with k-means for robust convergence.

In the final partitioning results, the majority of clusters were identified as clean clusters, while a minority of clusters were marked as backdoor samples. This fully unsupervised method requires no prior knowledge of the poisoning ratio and can be generalized to both datasets and model-level attacks. Since the method is based solely on post-measurement statistics, it is independent of any specific QNN architecture.

\section{QSentry: A Defense Framework Based on Measurement Clustering}

\begin{figure*}[t]
  \centering
  \includegraphics[width=\linewidth]{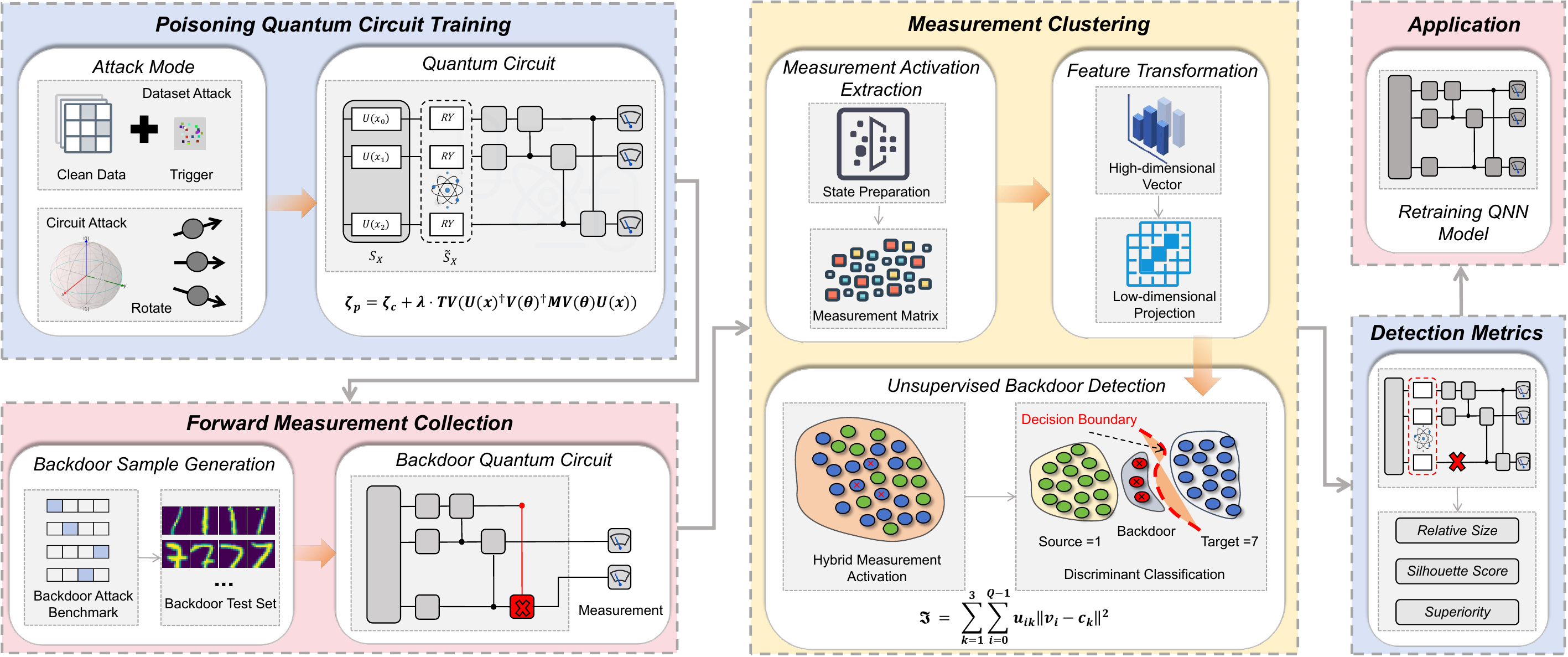}
  \caption{QSentry defense framework architecture. The system integrates poisoning training, forward measurement collection and measurement clustering.}
  \label{fig:qsentry_framework}
\end{figure*}

This section presents QSentry, a unified framework for detecting backdoor attacks in QNNs. As illustrated in Figure~\ref{fig:qsentry_framework}, QSentry integrates a post-training anomaly backdoor detection module based on Measurement Clustering, designed to identify both dataset and model-level backdoors without prior knowledge of the trigger pattern or poisoning strategy.

\subsection{Threat Model}

The threats posed to QNNs by backdoor attacks emanate from two primary attack vectors. The vectors in question represent distinct attack surfaces. One vector operates at the classical datasets, while the other manipulates the quantum circuit substrate directly. Both vectors have the capacity to embed hidden malicious functionality while preserving the model's performance on legitimate inputs.

\subsubsection{Data Poisoning Attack} In this scenario, an adversary injects a trigger pattern $\tau$ into a subset of training samples, generating backdoor inputs $x^\tau = T(x;\tau)$ and assigning them a target label $y^{t}$. The resulting poisoned dataset causes the model to learn a spurious correlation between the trigger and $y^t{}$, while maintaining high accuracy on clean inputs to avoid suspicion.

\begin{align}
    D_{\text{poison}} = \left \{ (x_{i}^{\tau}, y^{t}) \right \},
\end{align}

\subsubsection{Quantum Circuit Attack} Following the QTrojan paradigm\cite{Chu2023QTrojan}, attackers can directly modify the parameterized circuitry of a QNN by inserting malicious gates after the quantum-encoded circuit. By adjusting the parameters, the output can be made to conform to a target state defined by the attacker. This modification directly embeds hidden backdoor behavior into the model's quantum architecture.

\subsubsection{Defender Assumptions}
QSentry operates under the assumption that the defender has access to the trained QNN parameters and a set of test samples that may contain backdoor inputs. The framework does not require knowledge of the poisoning rate, trigger characteristics, or adversary capabilities, making it applicable to real-world QML deployment settings.

\subsection{Defense Goal}

The objective of QSentry is to detect backdoor attack anomalies in QNNs by identifying the abnormal disturbance they impose on qubit measurement statistics. These perturbations are both measurable and largely independent of the input class, enabling our defense to isolate backdoor samples directly in a transformed measurement space. Once identified, such samples can be removed. QSentry is attack-agnostic and requires no surrogate models, auxiliary classifiers, or access to internal quantum states.

\subsection{QSentry System Defense Architecture}

QSentry's operational framework is characterized by a three-stage pipeline, as illustrated in Figure~\ref{fig:qsentry_framework}. The objective of the design is to detect backdoors directly from observable quantum measurements. The workflow comprises: (1) \emph{Poisoning Quantum Circuit Training}, which simulates adversarial conditions to create real threat models. (2) \emph{Forward Measurement Collection} is used to collect measurement data from the quantum measurement layer, performed by the QNN. (3) \emph{Measurement Clustering}, which identifies backdoor samples through statistical deviations. The first stage itself is not part of the defense, but is used to reproduce a realistic threat model in order to evaluate the detector.

\subsection{Detailed Workflow \label{detailed_ workflow}}

\subsubsection{Poisoning Quantum Circuit Training}

To evaluate QSentry under trusted threat conditions, we first train a QNN exhibiting backdoor behavior. For dataset-level attacks, the model is trained using a hybrid dataset containing both clean and backdoor samples. Training begins with a clean initialization to avoid bias and applies standard gradient-based optimization with cross-entropy loss. The resulting model achieves high accuracy on clean inputs while associating trigger patterns with target labels, thus exhibiting typical backdoor characteristics (i.e., high clean accuracy and high attack success rate).

The training dataset is defined as:
\[
D = D_{\text{clean}} \cup D_{\text{poison}},
\]

where \(D_{\text{clean}}\) denotes clean samples without triggers, and \(D_{\text{poison}}\) denotes backdoor samples containing trigger patterns mapped to target labels. Typically, \(|D_{\text{clean}}| \gg |D_{\text{poison}}|\) to maintain high accuracy.

For clean samples, the cross-entropy loss is:
\begin{equation}
\zeta_c(x, y; \Theta) = -\log P(y \mid x; \Theta),
\end{equation}
where \(P(y \mid x; \Theta)\) denotes the predicted probability for label \(y\).  
For backdoor samples, the corresponding loss is:
\begin{equation}
\zeta_p(x^\tau, y^{t}; \Theta) = -\log P(y^{t} \mid x^\tau; \Theta),
\end{equation}
where \(x^\tau\) represents the backdoor of input \(x\) associated with a malicious target label \(y^{t} \neq y\).

Parameter updates are performed via gradient descent:
\begin{equation}
\Theta \leftarrow \Theta - \eta \nabla_\Theta (\zeta_c + \zeta_p),
\end{equation}
where \(\eta\) is the learning rate and \(\Theta\) denotes the trainable parameters. This dual-objective optimization embeds backdoor functionality while maintaining high clean accuracy.

For \textit{model-level} attacks, we adopt the QTrojan~\cite{Chu2023QTrojan} paradigm, inserting malicious parameterized gates after the state encoding layer. The parameters of these gates are optimized to maximize the overlap between the output state of the triggered input and the target state while preserving the performance on clean inputs.

The training employs a composite loss function:
\begin{align}
    \zeta(\Theta) = \frac{1}{|D|} [\sum_{i}^{N_{c}}  \zeta_c(x_{i}, y_{i}; \Theta) + \sum_{j}^{N_{p}}  \zeta_p(x^{\tau}_{j}, y_{j}^{t}; \Theta)],
\end{align}

where \(\zeta(\Theta)\) denotes the composite loss function for model training, \(|D|\) represents the total number of samples in the training batch \(D\), \(N_c\) is the number of samples in the clean sample subset \(D_{\text{clean}}\). \(N_p\) is the number of samples in the backdoor sample subset \(D_{\text{poison}}\). The poisoned loss includes a regularization term:
\begin{equation}
\zeta_p(x^{\tau}, y^{t}; \Theta) = 
\zeta_c(x, y ; \Theta) 
+ \lambda \cdot TV\!\left(U(x)^\dagger V(\Theta)^\dagger M V(\Theta) U(x)\right),
\end{equation}
where \(\lambda\) is a regularization coefficient and \(TV(\cdot)\) denotes the total variation term used to suppress anomalous measurement distributions caused by triggers. Here, \(U(x)\) represents the state encoding unitary, \(V(\Theta)\) the parameterized (possibly malicious) variational circuit, and \(M\) the measurement operator. This regularization ensures the backdoor remains concealed while maintaining clean accuracy.

This regularization suppresses anomalous measurement distributions caused by triggering, ensuring the backdoor remains concealed. The resulting model achieves high accuracy and attack success rate, establishing a reliable benchmark for detection experiments.

\subsubsection{Forward Measurement Collection}

Given a trained backdoored QNN and a set of test inputs, QSentry performs forward propagation and records the measurement outcomes of each qubit. These outputs form a measurement activation matrix
\[
\mathbf{A} \in \mathbb{R}^{M \times Q},
\]
where each row corresponds to one input sample, and each column corresponds to a qubit measurement outcome expectation value. Here, \(M\) denotes the number of test samples and \(Q\) the number of qubits. Due to the no-cloning theorem, the final measurement represents the only observable in the computation within the QNN, thus ensuring that our defenses are entirely based on the obtained data.

\begin{algorithm}[t]
\caption{QSentry Backdoor Detection Framework Workflow}
\label{alg:framework_algorithm}
\begin{algorithmic}[1]
\REQUIRE Trained QNN $f_\theta$, clean dataset $D_c$, test set $D_t$, backdoor dataset $D_p$
\ENSURE Detected backdoor samples in anomalous cluster $\mathbf{c}_{\text{anom}}$

\vspace{3pt}
\STATE \textbf{Step 1: Poisoning Training}
\STATE $w \gets w_0$ \COMMENT{Initialize with clean model}
\IF{attack type = model-level}
    \STATE $\theta = 2 \cdot \arccos(|\langle \psi_0 | \psi_1 \rangle|)$
    \STATE $|\psi_0\rangle \gets$ state after encoding
    \STATE $|\psi_1\rangle \gets$ target state
\ENDIF
\REPEAT
    \STATE $(x, y) \gets$ random sample from $D_c \cup D_p$
    \IF{$(x, y) \in D_p$}
        \IF{attack type = model-level}
            \STATE Add $R_Y(\theta)$ after each qubit encoding layer
        \ENDIF
        \STATE $w_p \gets \arg\min_w \zeta_p(x^p, y, w)$
    \ELSE
        \STATE $w_c \gets \arg\min_w \zeta_c(x, y, w)$
    \ENDIF
\UNTIL{convergence}

\vspace{3pt}
\STATE \textbf{Step 2: Measurement Collection}
\FOR{each sample $x$ in $D_t$}
    \STATE Run forward pass through $f_\theta$.
    \STATE Record qubit measurement vector $m_i$.
\ENDFOR
\STATE Construct measurement activation matrix $\mathbf{A} = [m_i]$.

\vspace{3pt}
\STATE \textbf{Step 3: Measurement Clustering}
\STATE $\mathbf{V} \leftarrow \text{reduce}(\mathbf{A})$ \COMMENT{dimension reduction}
\STATE $\mathcal{C} \leftarrow \text{cluster}(\mathbf{v})$ \COMMENT{unsupervised partitioning}
\STATE Identify anomalous cluster $\mathbf{c}_{\text{anom}}$

\vspace{3pt}
\STATE \textbf{Output:} Return $\mathcal{C}_{\text{anom}}$ as detected backdoor samples

\end{algorithmic}
\end{algorithm}

\subsubsection{Measurement Clustering for Anomaly Discovery}

This detection step employs unsupervised clustering of the quantum measurement matrix to identify anomalies caused by backdoors. By analyzing the reduced feature space, a compact feature representation is obtained that amplifies systematic deviations introduced by the backdoor. Next, clustering (e.g., K-Means~\cite{k_means}) is applied to partition the samples. A minority cluster exhibiting anomalous geometry and measurement statistics is identified as the backdoor cluster. The method isolates minority clusters that display deviant measurement statistics and geometric properties, which are characteristic of backdoor samples. This approach is attack-agnostic, requires no prior knowledge of the trigger pattern or circuit internals, and operates directly on measurement outcomes.

Using the Measurement Clustering method described in Chapter~\ref{Chapter_Measurement_Clustering}, quantum measurement statistics are transformed into discriminative features for backdoor identification. The detection process begins with quantum forward propagation, where each input sample \(x_i\) passes through the QNN to produce an output quantum state \(|\psi^{(i)}(\theta)\rangle\). Quantum measurements are then performed to obtain the expectation value of each qubit’s measurement observable:
\begin{equation}
    m(i, q) = \langle \psi^{(i)}(\theta) | M_q | \psi^{(i)}(\theta) \rangle,
\end{equation}

These expectation values form the measurement activation matrix \(\mathbf{A} \in \mathbb{R}^{M \times Q}\), where each row corresponds to one test sample and each column represents the expectation value of a qubit measurement. This matrix captures the statistical behavior of the QNN under different inputs and serves as the basis for subsequent clustering analysis. The high-dimensional matrix then undergoes dimensionality reduction via ICA:
\begin{equation}
    \mathbf{V}_{\text{reduced}} = \mathbf{A} \mathbf{W}^*
\end{equation}
where \(\mathbf{W}^*\) is the ICA-derived projection matrix and clustering is performed via the K-Means algorithm:
\begin{equation}
\{\mathbf{c}_{\text{clean}}, \mathbf{c}_{\text{backdoor}}\} 
= \text{K-Means}(\mathbf{V}_{\text{reduced}}),
\end{equation}
where \(\mathbf{c}_{\text{backdoor}}\) typically forms a minority cluster characterized by distinct measurement statistics. This unsupervised approach is attack-agnostic, requires no prior knowledge of the trigger or circuit internals, and operates directly on measurement outcomes.

\subsection{Algorithm Description}

To complement the system workflow described in Section~\ref{detailed_ workflow}, we provide a formal specification of the QSentry detection process. Algorithm~\ref{alg:framework_algorithm} summarizes the complete procedure, including quantum measurement extraction, forward measurement collection, and measurement clustering. This algorithm embodies the minimal-assumption principle of the framework and serves as a reproducible template for practical deployment.

Detection will be triggered when all preset separation metrics exceed their thresholds and a few clusters exhibit characteristics consistent with the expected backdoor. The integration of classical clustering techniques with quantum specificity measurement analysis ensures the method's scalability and adaptability across diverse attack scenarios, establishing Measurement Clustering as a reliable tool for post-training security evaluation of QML systems.

\section{Experimental Verification of Backdoor Defense}
\subsection{Experiment Setup}
\subsubsection{Dataset} The MNIST\cite{cohen2017emnistextension_mnisthandwritten} handwritten digit dataset is a classic dataset widely used in the field of DL and has become a standard benchmark for classification tasks. This dataset contains 60,000 training samples and 10,000 test samples. Each digit is stored as a 28×28 pixel grayscale image. For our experiments, we utilize a subset of MNIST tailored for binary classification (denoted as MNIST[1,7]). The selection of the dataset and the corresponding classification tasks took into account both the limitations of currently available qubits and the classification task settings widely used in the field of QML.

\subsubsection{QNN Models} Quantum Circuit Learning (QCL)\cite{Mitarai_2018_QCL} is a classical-quantum hybrid framework that utilizes QNNs. It improves performance on tasks such as high-dimensional regression or classification by non-linearly encoding classical data and using shallow, trainable variational circuits. In this work, we implement an 8-qubit circuit with a depth of eight layers. Each layer is composed of parameterized single-qubit rotations ($R_X, R_Z, R_X$) followed by a circular entangling gate. The output of the quantum circuit is measured by the expected values of $\langle Z_0\rangle$ and $\langle Z_1\rangle$ input to the Softmax function.

\subsubsection{Data Encoding} The operation of QNNs on classical data requires a preliminary step of encoding the data into quantum states. This quantum state encoding functions as a feature map that transforms data from $\mathbb{R}^Q$ into a state within a $2^q$-dimensional Hilbert space ($\mathbb{C}^{2^q}$). To enable efficient classical simulation of more qubits, we adopt amplitude encoding in this work. This method encodes a classical vector ${\bf{x}} \in \mathbb{R}^N$ into the amplitudes of a quantum state over $\lceil \log_{2}N \rceil$ qubits, formalized as $|{\bf{x}}\rangle = \sum_{i=1}^N x_i |i\rangle$, where ${|i\rangle}$ denotes the computational basis. A fundamental prerequisite for this encoding is that the input data must be normalized, satisfying $|{\bf{x}} |^2=\sum_i|x_i|^2=1$.

In the MNIST dataset, each image is a 28×28 pixel matrix. Direct amplitude encoding of such raw images requires 1024 amplitude values, equivalent to a 10-qubit system. This increases computational overhead and noise sensitivity, thus degrading performance. Therefore, we introduce a preprocessing step to crop the image size to 16×16 while minimizing the loss of key data features. After this processing, encoding requires only 256 amplitude values in an 8-qubit system, thereby reducing resource requirements and noise impact while maintaining feature validity.

\subsubsection{Backdoor Samples Generation} 
\begin{figure}[htbp]
    \centering
    \includegraphics[width=\columnwidth]{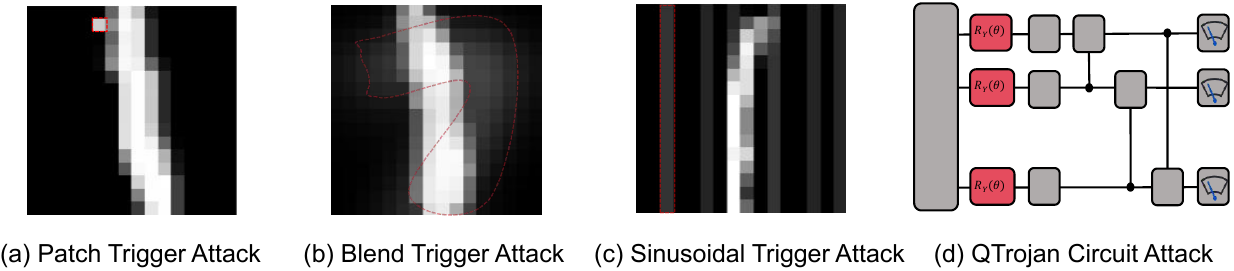}
    \caption{Examples of different backdoor attacks. (a) Patch Trigger Attack, (b) Blend Trigger Attack, and (c) Sinusoidal Trigger Attack are three-data attacks, (d) QTrojan Circuit Attack is the quantum model-level attack.}
    \label{fig:Four_Attack_Type}
\end{figure}

In the experimental setup for backdoor sample generation, we evaluated four representative strategies, all of which follow a unified paradigm of fixed target labeling and location injection, as shown in Figure~\ref{fig:Four_Attack_Type}. Specifically, these include: (a) \textbf{Patch Trigger Attacks}, which randomly inject pixel block perturbations into high-contrast color regions. (b) \textbf{Blend Trigger Attack}, where target class samples are blended with the original samples using coefficients $\sigma = 1.5$ and $\lambda = 0.3$ to achieve progressive contamination in the frequency domain. (c) \textbf{Sinusoidal Trigger Attack}, which overlays a Gaussian-filtered standard trigger template with a mixing coefficient defined by angle 0.2 and frequency 1. (d) \textbf{QTrojan Circuit Attack}, where a parameterized malicious subcircuit is implanted after the quantum encoding layer, constructing the trigger using rotated $R_Y(\theta)$ gates. Backdoor samples for each attack type are generated independently based on different poisoning rates.

\subsubsection{Evaluation Metrics}
To comprehensively assess both the effectiveness of backdoor implantation and the performance of detection mechanisms, the experiments employ a set of evaluation metrics.

\textbf{Clean Accuracy (CA)} measures the classification performance of a poisoned model on the original test set without implanted triggers, thus reflecting the stealth of the attack. Its definition is as follows:
\begin{align}
    \text{CA}=\frac{1}{N_{\text{clean}}}\sum_{i=1}^{N_{\text{clean}}}\mathbb{I}\bigl(\hat{y}_i=y_i\bigr),
\end{align}
where $N_{\text{clean}}$ is the number of clean test samples, and $\hat{y}_i$ and $y_i$ denote the predicted and true labels, respectively.

\textbf{Attack Success Rate (ASR) }measures the effectiveness of a model in predicting the attacker's target label on test samples containing the trigger after the trigger has been successfully implanted. It is defined as follows:
\begin{align}
    \text{ASR}=\frac{1}{N_{\text{backdoor}}}\sum_{i=1}^{N_{\text{backdoor}}}\mathbb{I}\bigl(\hat{y}_i=y_{\text{target}}\bigr),
\end{align}
where $N_{\text{backdoor}}$ is the number of backdoor test samples and $y_{\text{target}}$ is the target-class label enforced by the attacker.

\textbf{Silhouette Coefficient (SC)} evaluates the geometric separability of clusters in the reduced measurement-feature space, providing an indicator of how distinguishable backdoor samples are from clean ones:
\begin{align}
    s(i)=\frac{b(i)-a(i)}{\max\{a(i),\, b(i)\}},
\end{align}
where $a(i)$ is the average intra-cluster distance of sample $i$, and $b(i)$ is the minimum average distance between sample $i$ and all points in any other cluster.  
The overall score is:
\begin{align}
    \text{SC}=\frac{1}{N}\sum_{i=1}^{M}s(i),
\end{align}
with $N$ being the total number of samples. Higher SC values indicate strong cluster separability.

\textbf{Relative Cluster Size (RCS)} reflects the proportion of samples assigned to the minority cluster, which is expected to correspond to backdoor samples. It measures whether the clustering result matches the theoretical poisoning ratio:
\begin{align}
    \text{RCS}=\frac{|\mathcal{C}_{\text{minority}}|}{|\mathcal{C}_{\text{clean}}| + |\mathcal{C}_{\text{backdoor}}|},
\end{align}
where $\mathcal{C}_{\text{minority}}$ denotes the cluster with the smaller number of samples.  
An RCS close to the true poisoning rate indicates consistent backdoor cluster structure.

\textbf{Detection Accuracy (DA)} measures how well the clustering-based detection method correctly distinguishes backdoor and clean samples:
\begin{align}
    \text{DA} = \frac{TP + TN}{TP + TN + FP + FN},
\end{align}
where $TP$, $TN$, $FP$, and $FN$ denote true positives, true negatives, false positives, and false negatives, respectively.

\textbf{ F1 score} captures the balance between correct identification of backdoor samples and reduction of false alarms, making it suitable for evaluating detection tasks with minority backdoor samples:
\begin{align}
    \text{F1} = 2 \cdot \frac{TP}{2 TP + FP + FN}.
\end{align}
A higher F1 score indicates that the detection method effectively identifies backdoor samples without overestimating false alarms.

We implemented QSentry using the PennyLane framework, which provides a unified platform for constructing and training both classical and quantum machine learning models. All evaluation tasks were performed on a workstation equipped with an Intel(R) Core(TM) i9-14900K CPU, an NVIDIA GeForce RTX 4090 GPU, and 125 GB of RAM.

\begin{figure}
  \centering
  \includegraphics[width=\linewidth]{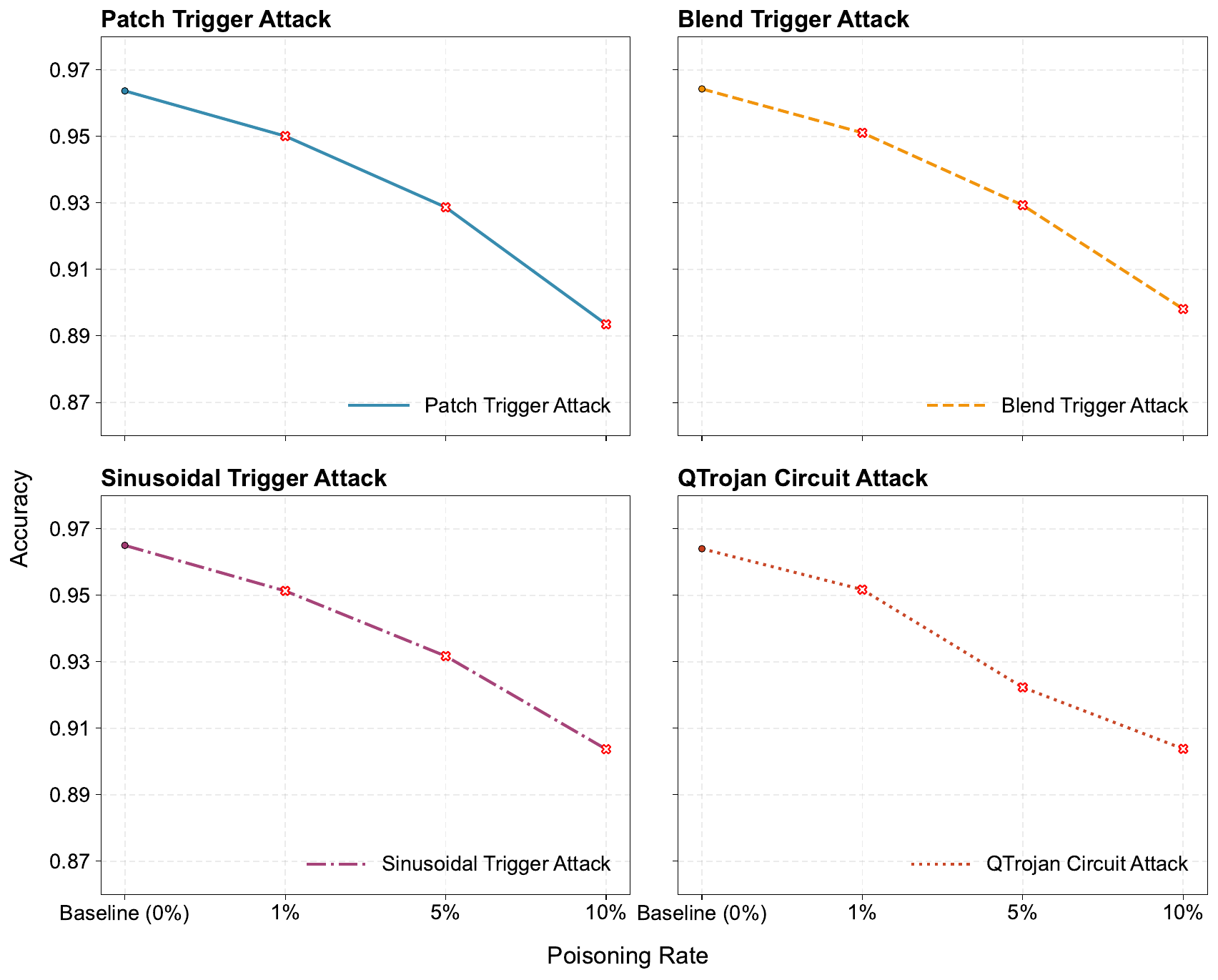} 
  \caption{Comparison of classification accuracy of the binary classification task MNIST[1,7] under four types of backdoor attacks and three poisoning rates.}
  \label{fig: Attacks}
\end{figure}

\subsection{Construction and Benchmark Evaluation of QNN Backdoor Attack Threat Model}
Research into security vulnerabilities in QML is still in its early stages. Backdoor attacks, as a covert and powerful threat, pose a serious risk to model integrity, and their potential risks have not been fully explored. To develop effective backdoor defense mechanisms, it is essential to construct and validate a realistic threat model that demonstrates the severity of such risks. If a defense method cannot withstand a real and effective attack, its evaluation results will lack persuasiveness. Therefore, this experiment systematically constructs a backdoor attack benchmark for QNNs. Our goal is to train a clean QNN model as a performance baseline and then attack this model using four different backdoor attack methods to verify the feasibility and effectiveness of backdoor attacks in QNNs, and to provide experimental evidence for subsequent defense research. All experiments are detailed and reproducible; experimental data were repeated 3-5 times and averaged to mitigate uncertainty.

The performance trends in Figure~\ref{fig: Attacks} clearly demonstrate the effectiveness and stealthiness of the backdoor threats. The baseline QNN achieves the highest accuracy on the clean dataset, indicating its inherent performance advantage. At a low poisoning rate, all four attacks only cause a slight decrease in CA, highlighting their stealthiness. As the poisoning rate increases to 5\% and 10\%, the differences in the impact of different attacks become more pronounced, with most attacks reducing the overall accuracy to a lower level. \textbf{Crucially, the model maintains a relatively high classification accuracy even under severe malicious input conditions, which further highlights the stealth of the attack and the model's stability with clean input data.}

\begin{figure}[t]
  \centering
  \includegraphics[width=\linewidth]{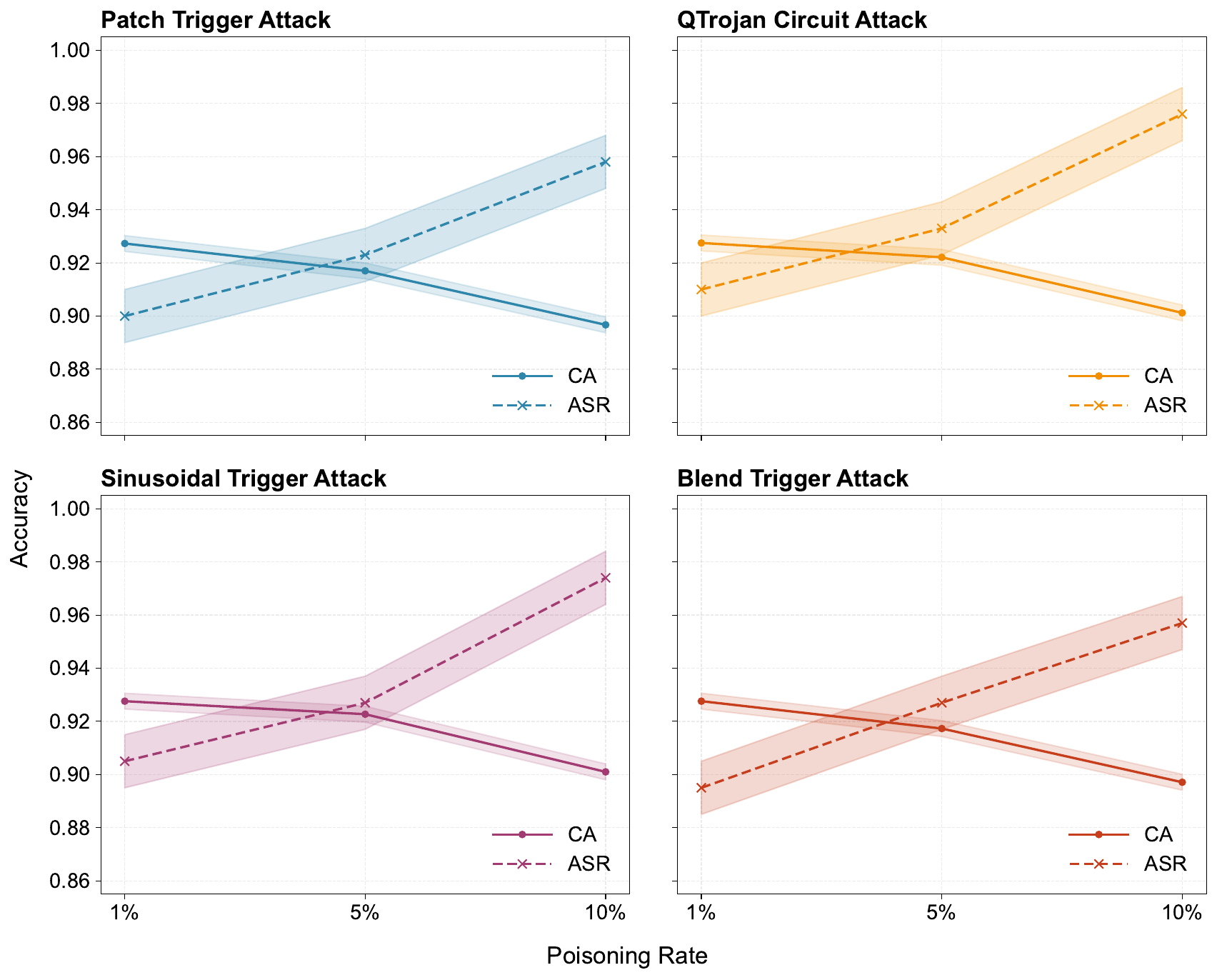}
  \caption{
  Variation of CA and ASR under different poisoning ratios for four representative backdoor attacks (Patch Trigger Attack, Blend Trigger Attack, Sinusoidal Trigger Attack, and QTrojan Circuit Attack). 
  }
  \label{fig:poisoned_attacks}
\end{figure}

To further evaluate the threat model, we assessed the CA on clean samples and the ASR on backdoor samples of the poisoned model under different poisoning rates for each type of backdoor attack.
As shown in Figure~\ref{fig:poisoned_attacks}, the horizontal axis represents the poisoning ratio applied to the training data, and the vertical axis represents the corresponding CA and ASR. When the poisoning ratio increases from 1\% to 10\%, the CA of all four attacks remains at a high level, while the ASR gradually increases with the increase in the poisoning ratio. The experimental results show that at different poisoning rates, the CA of all attacks remains at a high level, indicating their strong stealthiness. The ASR increases rapidly with the increase in the poisoning ratio and stabilizes at a high level, indicating that backdoors can be effectively implanted and triggered without affecting the stability of the model.

This experiment systematically evaluates the effectiveness and stealth of four representative backdoor attacks on the employed QNN architecture. Experimental findings demonstrate that while maintaining high accuracy on clean test samples, most attacks achieve a high ASR, and the ASR stabilizes as training converges. This outcome underscores the practical significance of the constructed backdoor threat scenario and highlights its potential to challenge the efficacy of future defense methodologies. In light of the findings from this experiment, the subsequent experiments in this paper will utilize these attacks as a baseline threat model. This approach will serve to verify the effectiveness of the proposed defense method.

\subsection{Comprehensive Performance Evaluation of QSentry}
This experiment evaluates the efficacy of our proposed QSentry defense framework against verified backdoor threats in QNNs. Under the scenario of a QNN trained on a contaminated dataset and a small, clean validation set, we compare QSentry with baseline methods. The objective is to validate its superior capability in extracting discriminative features from quantum states compared to classical raw-data features and to demonstrate its general defensive performance across diverse backdoor attacks.

\begin{table*}[t]
\centering
\caption{
Detection performance of QSentry versus Raw Clustering under four attack modes and three poisoning rates. Bold entries indicate cases where QSentry outperforms the baseline on the corresponding metric.}
\label{table:poison-detection}
\renewcommand{\arraystretch}{1.15}
\setlength{\tabcolsep}{8pt}
\begin{tabular}{lcccccc}
\toprule
& \multicolumn{2}{c}{\textbf{1\% Poison Rate}}
& \multicolumn{2}{c}{\textbf{5\% Poison Rate}}
& \multicolumn{2}{c}{\textbf{10\% Poison Rate}} \\
\cmidrule(lr){2-3}
\cmidrule(lr){4-5}
\cmidrule(lr){6-7}
\textbf{Attack Mode} 
& \textbf{QSentry} & \textbf{Raw}
& \textbf{QSentry} & \textbf{Raw}
& \textbf{QSentry} & \textbf{Raw} \\
\midrule

Patch Trigger 
& 99.8\% / \textbf{83.3\%} & 98.4\% / 61.4\%
& 99.5\% / \textbf{90.9\%} & 97.9\% / 70.3\%
& 99.7\% / \textbf{97.1\%} & 97.8\% / 78.4\% \\

Blend Trigger
& 99.7\% / \textbf{76.9\%} & 97.9\% / 64.1\%
& 99.4\% / \textbf{89.3\%} & 98.4\% / 70.4\%
& 99.7\% / \textbf{97.1\%} & 98.9\% / 79.1\% \\

Sinusoidal Trigger
& 99.5\% / \textbf{71.4\%} & 97.8\% / 62.7\%
& 98.9\% / \textbf{82.0\%} & 96.7\% / 70.0\%
& 98.9\% / \textbf{90.1\%} & 95.8\% / 78.9\% \\

QTrojan Circuit
& 99.6\% / \textbf{71.4\%} & 98.4\% / 62.3\%
& 98.8\% / \textbf{80.6\%} & 96.1\% / 71.6\%
& 98.7\% / \textbf{88.5\%} & 94.6\% / 78.1\% \\

\bottomrule
\end{tabular}
\end{table*}

The quantitative evaluation of detection performance is summarised in Table~\ref{table:poison-detection}. The results report both detection accuracy and F1 score across four representative backdoor attack modes and three poisoning rates. In the comparison, QSentry represents our proposed measurement-based defense framework, while Raw represents the baseline method that directly applies clustering to the raw pixel domain without utilizing quantum measurement information.

Across all attack settings, QSentry consistently and substantially outperforms the Raw baseline in terms of both accuracy and F1 score. It is worth noting that, because backdoor samples constitute an extremely small fraction of the dataset, accuracy alone becomes an unreliable indicator of detection performance—it may remain high even when backdoor samples are poorly identified. In contrast, the F1 score offers a more faithful characterization of defense effectiveness under severe class imbalance, capturing the essential trade-off between precision and recall. The significant improvement in the F1 score clearly demonstrates that backdoor-triggered samples exhibit unique and statistically separable activation characteristics in the quantum measurement space. This strongly suggests that backdoors are more easily detected via QSentry.

Furthermore, QSentry demonstrates strong robustness in detecting more stealthy backdoor strategies, including frequency-domain triggers and adaptive attacks crafted to exploit quantum circuit characteristics. \textbf{Overall, the empirical results validate QSentry as a general and effective defense mechanism for QNNs, capable of accurately isolating backdoor samples while maintaining high precision and high F1 scores}. Its performance consistently surpasses that of pixel-space baselines, underscoring the advantage of leveraging QSentry for reliable backdoor detection.

\subsection{Backdoor Detection Verification Based on Silhouette Coefficient}

\begin{figure}[t]  
    \centering
    \includegraphics[width=\linewidth]{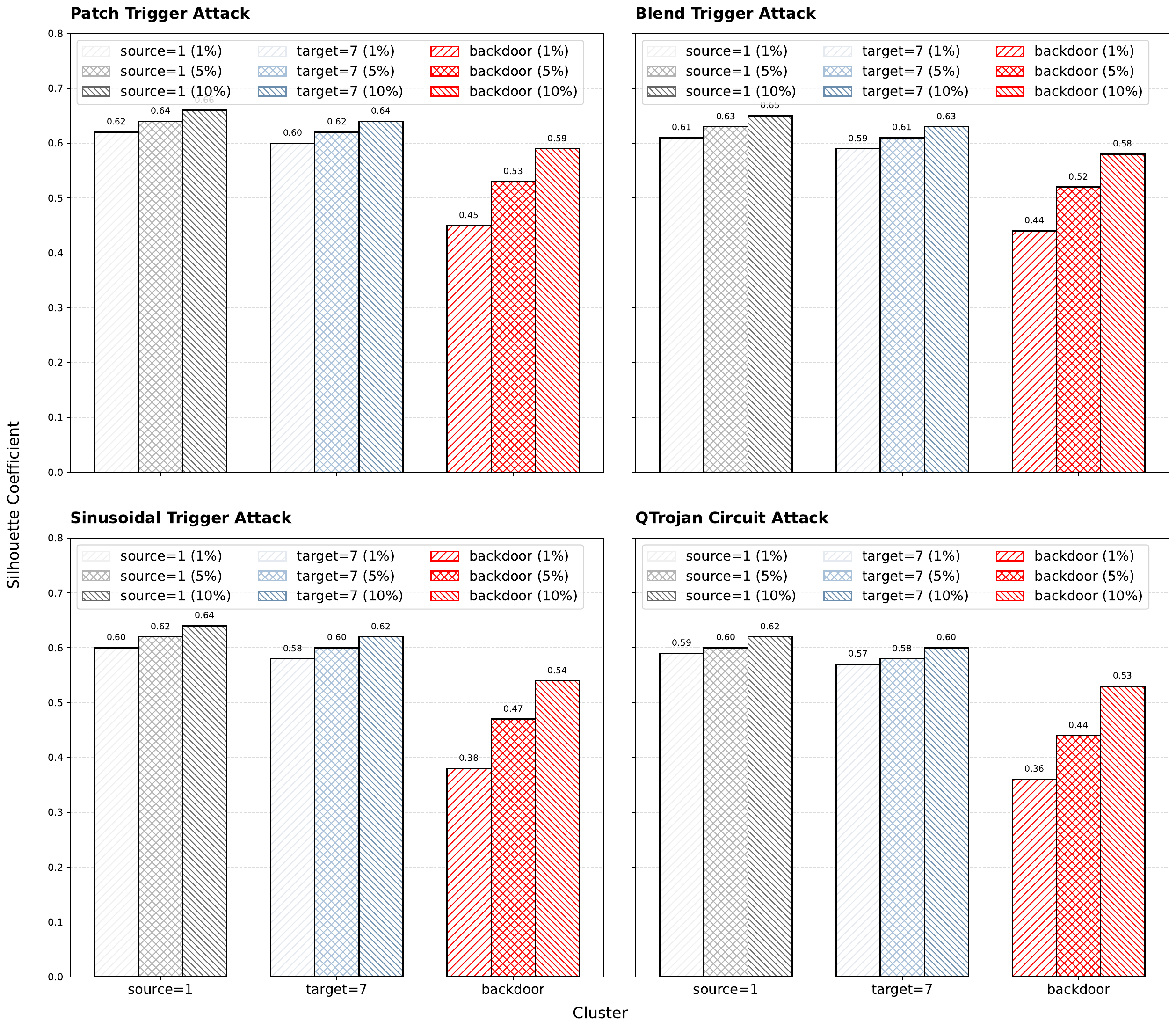}
    \caption{Silhouette-based clustering analysis across four backdoor attack types—Patch Trigger, Blend Trigger, Sinusoidal Trigger, and QTrojan Circuit Attack—under poisoning rates of 1\%, 5\%, and 10\%.}
    \label{fig: Silhouette}
\end{figure}

This experiment aims to quantitatively verify the effectiveness and robustness of the QSentry defense framework in detecting backdoor attacks. We hypothesize that poisoned data trigger a bicluster distribution in the measurement space of the quantum model, while clean data conform to a monocluster distribution. To verify this hypothesis, we introduce the silhouette coefficient as the core indicator for quantifying distribution separation. By calculating this coefficient and comparing it with a predefined threshold, we systematically evaluate the defense mechanism's ability to distinguish between malicious and clean samples under various attack scenarios.

Across all four attack modes and poisoning ratios, clean source and target samples consistently maintain high silhouette values, indicating that benign measurement activations naturally form compact and coherent clusters within the quantum feature space. In contrast, backdoor samples exhibit substantially lower silhouette scores at all poisoning levels, with values dropping to the 0.36–0.45 range under a 1\% poisoning rate. As illustrated in Figure~\ref{fig: Silhouette}, these degraded scores confirm that backdoor triggers introduce identifiable geometric distortions to the measurement distribution, even when the contamination is extremely sparse.

As the poisoning rate increases to 5\% and 10\%, the silhouette coefficients increase moderately, reflecting that a larger proportion of backdoor inputs yields a more structurally coherent cluster. This trend is consistent with theoretical expectations: higher poisoning density produces stronger activation biases, amplifying the backdoor signature and improving cluster consistency within the backdoor subset. Importantly, this pattern remains consistent across visually obvious triggers, subtle frequency-domain triggers, and model-level manipulations, demonstrating that the quantum measurement space preserves separability against both dataset and model-level threats.

Overall, the silhouette-coefficient results validate the core hypothesis behind QSentry: \textbf{backdoor attacks imprint statistically distinct signatures on quantum measurement activations}. These signatures manifest as minority clusters that are reliably distinguishable from clean data. The consistency of this phenomenon across poisoning intensities and attack modalities confirms the robustness, generality, and practical viability of measurement-based anomaly detection in QNNs.

\subsection{Backdoor Detection Verification Based on Cluster Relative Size}

\begin{table*}[t]
\centering
\caption{Actual (Act) versus Predicted (Pred) QSentry clustering counts (total 1000 samples).
         $\Delta = \mathrm{Pred} - \mathrm{Act}$.}
\label{tab:QSentry-cluster-ideal-vs-pred}
\renewcommand{\arraystretch}{1.15}
\footnotesize

\begin{tabular}{cc | ccc| ccc | ccc | ccc}
\toprule
\textbf{Poison Rate} & \textbf{Category} 
& \multicolumn{3}{p{2.6cm}|}{\centering\textbf{Patch Trigger Attack}}
& \multicolumn{3}{p{2.7cm}|}{\centering\textbf{Blend Trigger Attack}}
& \multicolumn{3}{p{3.2cm}|}{\centering\textbf{Sinusoidal Trigger Attack}}
& \multicolumn{3}{p{2.9cm}}{\centering\textbf{QTrojan Circuit Attack}} \\
\cmidrule(lr){3-5} \cmidrule(lr){6-8} \cmidrule(lr){9-11} \cmidrule(lr){12-14}
& 
& \textbf{Act} & \textbf{Pred} & \textbf{$\Delta$}
& \textbf{Act} & \textbf{Pred} & \textbf{$\Delta$}
& \textbf{Act} & \textbf{Pred} & \textbf{$\Delta$}
& \textbf{Act} & \textbf{Pred} & \textbf{$\Delta$} \\
\midrule

\multirow{3}{*}{1\%} 
& Source 1 & $495$ & \textbf{$495$} & $0$   & $495$ & \textbf{$494$} & $-1$ & $495$ & \textbf{$495$} & $0$ & $495$ & \textbf{$493$} & $-2$ \\
& Target 7 & $500$ & \textbf{$498$} & $-2$  & $500$ & \textbf{$498$} & $-2$ & $500$ & \textbf{$496$} & $-4$ & $500$ & \textbf{$498$} & $-2$ \\
& Backdoor &   $5$ &   \textbf{$7$} & $+2$  &   $5$ &   \textbf{$8$} & $+3$ &   $5$ &   \textbf{$9$} & $+4$ &   $5$ &   \textbf{$9$} & $+4$ \\
\midrule

\multirow{3}{*}{5\%} 
& Source 1 & $475$ & \textbf{$475$} & $0$ & $475$ & \textbf{$474$} & $-1$ & $475$ & \textbf{$473$} & $-2$ & $475$ & \textbf{$472$} & $-3$ \\
& Target 7 & $500$ & \textbf{$495$} & $-5$ & $500$ & \textbf{$495$} & $-5$ & $500$ & \textbf{$491$} & $-9$ & $500$ & \textbf{$491$} & $-9$ \\
& Backdoor &  $25$ &  \textbf{$30$} &  $+5$ &  $25$ &  \textbf{$31$} &  $+6$ &  $25$ &  \textbf{$36$} & $+11$ &  $25$ &  \textbf{$37$} & $+12$ \\
\midrule

\multirow{3}{*}{10\%} 
& Source 1 & $450$ & \textbf{$450$} & $0$ & $450$ & \textbf{$450$} & $0$ & $450$ & \textbf{$449$} & $-1$ & $450$ & \textbf{$446$} & $-4$ \\
& Target 7 & $500$ & \textbf{$497$} & $-3$ & $500$ & \textbf{$497$} & $-3$ & $500$ & \textbf{$490$} & $-10$ & $500$ & \textbf{$491$} & $-9$ \\
& Backdoor & $ 50$ &  \textbf{$53$} & $+3$ &  $50$ &  \textbf{$53$} &  $+3$ &  $50$ &  \textbf{$61$} & $+11$ &  $50$ &  \textbf{$63$} & $+13$ \\
\bottomrule

\end{tabular}
\end{table*}

We found that when QNNs process data containing backdoor samples, the activation distribution in their measurement space typically exhibits a distinct bi-cluster structure. Further analysis revealed that~\ref{fig: Visual_Analysis}, unlike the primary clusters formed by uncontaminated samples, backdoor samples usually aggregate into smaller secondary clusters. This phenomenon demonstrates relative stability under various attack methods.

This study proposes using the relative size of clusters as a standard for detecting backdoor samples to realize a backdoor detection mechanism in a quantum model. The objectives of this experiment are twofold: first, to verify whether backdoor samples consistently exhibit a relatively small cluster structure under different backdoor attack types and poisoning rates; and second, to evaluate whether setting a cluster size threshold close to a preset poisoning rate allows for stable identification and localization of backdoor clusters without requiring additional manipulation of the model structure or parameters. To this end, we systematically experimented to verify this criterion in terms of geometric separability and numerical consistency, combining visualization analysis and cluster statistics methods.

\begin{figure}[t] 
    \includegraphics[width=\linewidth]{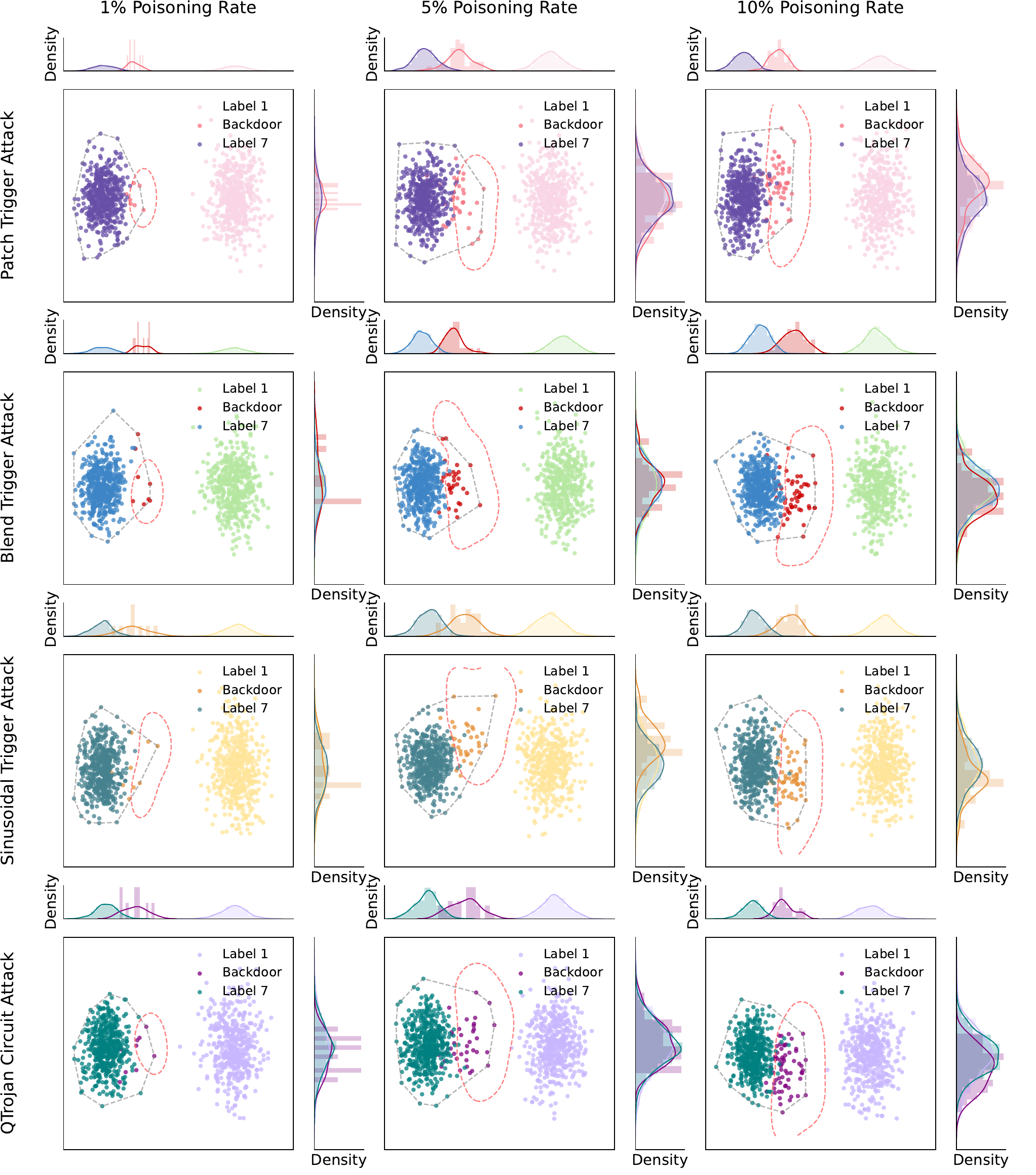}
    \caption{Visualization of different clusters. Visual analysis of the sample distribution under the four attack methods and three infection rates reveals that the Backdoor sample cluster consistently exhibits spatially distinguishable characteristics from the two clean sample clusters.}
    \label{fig:cluster_visualization}
\end{figure}

The experimental results demonstrate a high degree of consistency between the cluster structure and the relative size statistics. As illustrated in Figure~\ref{fig:cluster_visualization}, irrespective of the employed attack method, the activation space of the measured values is organized into three distinct clusters: two primary clusters comprising clean samples from the source and target classes, which collectively account for a substantial proportion; and a secondary cluster consisting of backdoor samples, which is comparatively diminutive in size and exhibits discernible boundary structures that delineate it from the primary clusters.

In addition, as demonstrated in Table~\ref{tab:QSentry-cluster-ideal-vs-pred}, the predicted cluster sizes exhibit a high degree of congruence with the preset poisoning rates. At a 10\% poisoning rate, the predicted backdoor cluster sizes are between 52 and 63, which is very close to the ideal value of 50. At 5\% and 1\% poisoning rates, the predicted backdoor cluster sizes are between 30 and 37 and 7 and 9, respectively, which is also quite consistent with the actual proportions. While there is a slight overestimation under certain attack methods, the overall deviation remains within an acceptable range and will not affect the reliable identification of backdoor sample clusters. Visualization and tabular data confirmed the stability of the structural feature that \textbf{backdoor samples cluster into small clusters, while pure samples cluster into large clusters} under all experimental conditions.

The results demonstrate that the relative cluster size is a reliable indicator for backdoor detection, as the minority clusters consistently correspond to backdoor samples and closely match the preset poisoning ratios. This behavior remains stable across different attack intensities. Overall, both visual clustering structure and quantitative statistics validate the feasibility and practicality of the QSentry method in quantum backdoor defense, providing a solid foundation for constructing secure QNN protection mechanisms.

\subsection{Comparative Analysis of QSentry Against Advanced Detection Methods}
To comprehensively evaluate the performance advantages of the QSentry defense framework, this experiment systematically compared it with three state-of-the-art backdoor detection methods: Identifying Backdoor Data with Optimized Scaled Prediction Consistency based on input consistency (MSPC)\cite{hou2024_ibdpsc}; Amplifying Anomalies in Backdoor Models through Knowledge Distillation (Distill to Detect)\cite{Distill_to_Detect}, and quantum-specific Q-Detection\cite{QDetection2025}. The evaluation aimed to quantify the detection capabilities of these four methods against various backdoor attack types in a unified test environment and verify the overall superiority of QSentry in terms of detection effectiveness.

\begin{table*}[t]
\centering
\caption{
Detection Accuracy / F1 score comparison among MSPC, Distill to Detect, Q-Detection, and QSentry under three poisoning rates.
Each cell reports ``Accuracy / F1''. Smaller gaps are used to highlight subtle performance differences.
}
\label{tab:comparison_poison_rates}

\begin{tabular}{@{}lcccc@{}}
\toprule
\textbf{Poisoning Rate: 1\%} 
& \textbf{MSPC} 
& \textbf{Distill to Detect} 
& \textbf{Q-Detection} 
& \textbf{QSentry (Ours)} \\ 
\midrule
Patch Triggered Attack      
& 99.2\% / 77.1\%  
& 99.0\% / 77.8\%  
& 99.6\% / 81.2\%  
& \textbf{99.8\% / 83.3\%} \\
Blend Triggered Attack      
& 99.0\% / 73.9\%  
& 98.8\% / 72.5\%  
& 99.5\% / 74.8\%  
& \textbf{99.7\% / 76.9\%} \\
Sinusoidal Triggered Attack 
& 98.8\% / 68.9\%  
& 98.6\% / 68.2\%  
& 99.3\% / 69.1\%  
& \textbf{99.5\% / 71.4\%} \\
QTrojan Circuit Attack      
& 98.9\% / 67.6\%  
& 98.7\% / 65.9\%  
& 99.4\% / 69.8\%  
& \textbf{99.6\% / 71.4\%} \\
\bottomrule
\end{tabular}

\vspace{8pt}

\begin{tabular}{@{}lcccc@{}}
\toprule
\textbf{Poisoning Rate: 5\%} 
& \textbf{MSPC} 
& \textbf{Distill to Detect} 
& \textbf{Q-Detection} 
& \textbf{QSentry (Ours)} \\ 
\midrule
Patch Triggered Attack      
& 98.9\% / 85.9\%  
& 98.7\% / 84.5\%  
& 99.3\% / 88.5\%  
& \textbf{99.5\% / 90.9\%} \\
Blend Triggered Attack      
& 98.7\% / 84.0\%  
& 98.5\% / 83.7\%  
& 99.2\% / 87.2\%  
& \textbf{99.4\% / 89.3\%} \\
Sinusoidal Triggered Attack 
& 98.1\% / 76.4\%  
& 97.9\% / 75.8\%  
& 98.7\% / 80.6\%  
& \textbf{98.9\% / 82.0\%} \\
QTrojan Circuit Attack      
& 97.9\% / 76.3\%  
& 97.7\% / 75.6\%  
& 98.6\% / 78.2\%  
& \textbf{98.8\% / 80.6\%} \\
\bottomrule
\end{tabular}

\vspace{8pt}

\begin{tabular}{@{}lcccc@{}}
\toprule
\textbf{Poisoning Rate: 10\%} 
& \textbf{MSPC} 
& \textbf{Distill to Detect} 
& \textbf{Q-Detection} 
& \textbf{QSentry (Ours)} \\ 
\midrule
Patch Triggered Attack      
& 99.1\% / 90.9\% 
& 98.9\% / 90.2\%  
& 99.4\% / 94.8\%  
& \textbf{99.7\% / 97.1\%} \\
Blend Triggered Attack      
& 99.0\% / 93.7\%  
& 98.7\% / 94.3\%  
& 99.3\% / 95.6\%  
& \textbf{99.7\% / 97.1\%} \\
Sinusoidal Triggered Attack 
& 97.8\% / 85.0\%  
& 97.5\% / 84.8\%  
& 98.4\% / 88.5\%  
& \textbf{98.9\% / 90.1\%} \\
QTrojan Circuit Attack      
& 97.4\% / 83.1\%  
& 97.2\% / 82.6\%  
& 98.2\% / 86.9\%  
& \textbf{98.7\% / 88.5\%} \\
\bottomrule
\end{tabular}

\end{table*}

Table~\ref{tab:comparison_poison_rates} provides a comprehensive comparison of the performance of four representative backdoor detection methods: MSPC, Distill to Detect, Q-Detection, and QSentry under four attack modes and three poisoning rates. QSentry maintains the highest F1 score and detection accuracy under all settings, demonstrating its superior ability to identify backdoor samples, even at extremely low poisoning rates. This improvement is especially notable at low poisoning levels, where traditional metrics like accuracy are unreliable due to class imbalance. In contrast, QSentry maintains a high F1 score, indicating robust detection sensitivity.

Q-Detection ranks second overall, closely following QSentry, performing well across all attack modes and poisoning levels. Its detection capability steadily improves with increasing poisoning rate, but remains slightly inferior to QSentry, suggesting that the QSentry method is more effective than Q-Detection's hybrid quantum-classical representation in capturing anomalies caused by backdoors in QNNs. The results for MSPC and Distill to Detect are similar, both outperforming Raw Clustering but lagging behind QSentry and Q-Detection. Their accuracy remains relatively high, but their F1 scores are significantly lower, especially under conditions of subtle or widely distributed trigger signals, indicating insufficient robustness in distinguishing between sparse backdoor and clean samples. The performance of both methods improves slightly with increasing poisoning rate, but the performance gap with QSentry remains significant.

Overall, the results confirm that \textbf{QSentry achieves the most balanced and reliable detection performance across all attack modes and poisoning intensities. It significantly improves the F1 score while maintaining high accuracy}, making it a practical and effective framework for backdoor threat detection in QML.

\section{Conclusion}
QSentry provides a practical and effective defense framework against backdoor threats in QNNs. Unlike methods that rely on inaccessible intermediate quantum states, QSentry utilizes measurement layer activation statistics, bypassing the fundamental limitation of quantum observability. Extensive experiments under various dataset and model-level attacks have confirmed that the quantum measurement distribution contains stable and detectable structural anomalies, enabling QSentry to accurately identify backdoor samples even with poisoning rates as low as 1\%. Compared to existing classical and hybrid defense methods, this framework consistently achieves higher detection precision.

Despite its strengths, QSentry has several limitations. First, its performance relies on a sufficient number of measurement samples, which can introduce significant runtime overhead on real devices. Additionally, the framework has primarily been evaluated on small-scale circuits and binary classification tasks, so its scalability to higher-dimensional feature spaces and deeper variational architectures remains to be verified. Finally, if attackers design adaptive triggers to evade detection by minimizing anomalies in the measurement space, this poses a challenge to current strategies.

\section{Ethics Considerations}
This study did not involve human subjects, personal data, or interactions with deployed systems. All experiments were conducted in a controlled environment, using publicly available datasets and simulated quantum circuits based on the PennyLane framework. Therefore, no ethical issues were identified.

\section{LLM Usage Considerations}
In the preparation of this manuscript, the authors utilized large language models (LLMs) such as ChatGPT for assistance with text polishing and grammatical correction only. The fundamental research ideas, formulation of the methodology, experimental design, execution, data interpretation, and scientific conclusions were solely and independently carried out by the authors.

\bibliographystyle{IEEEtran}
\bibliography{References}

@article{PQC,
  title={Parameterized quantum circuits as machine learning models},
  author={Benedetti, Marcello and Lloyd, Erika and Sack, Stefan and Fiorentini, Mattia},
  journal={Quantum Science and Technology},
  volume={4},
  number={4},
  pages={043001},
  year={2019},
  doi={10.1088/2058-9565/ab4eb5},
  publisher={IOP Publishing}
}

@article{Classify,
  author={Schuld, Maria and Bocharov, Alex and Svore, Krysta M. and Wiebe, Nathan},
  title={Circuit-centric quantum classifiers},
  journal={Physical Review A},
  volume={101},
  number={3},
  pages={032308},
  year={2020},
  doi={10.1103/PhysRevA.101.032308}
}

@article{Quantum_Modeling,
  author={Lloyd, Seth and Weedbrook, Christian},
  title={Quantum Generative Adversarial Learning},
  journal={Physical Review Letters},
  volume={121},
  number={4},
  pages={040502},
  year={2018},
  doi={10.1103/PhysRevLett.121.040502}
}

@article{Quantum_Chemistry_Simulation,
  author={Peruzzo, Alberto and McClean, Jarrod and Shadbolt, Peter and Yung, Man-Hong and Zhou, Xiao-Qi and Love, Peter J. and Aspuru-Guzik, Alán and O'Brien, Jeremy L.},
  title={A variational eigenvalue solver on a photonic quantum processor},
  journal={Nature Communications},
  volume={5},
  pages={4213},
  year={2014},
  doi={10.1038/ncomms5213}
}

@article{Biamonte_2017_qml,
   title={Quantum machine learning},
   volume={549},
   ISSN={1476-4687},
   url={http://dx.doi.org/10.1038/nature23474},
   DOI={10.1038/nature23474},
   number={7671},
   journal={Nature},
   publisher={Springer Science and Business Media LLC},
   author={Biamonte, Jacob and Wittek, Peter and Pancotti, Nicola and Rebentrost, Patrick and Wiebe, Nathan and Lloyd, Seth},
   year={2017},
   month=sep, pages={195–202} }

@article{shi2025qsan,
  title={QSAN: A Near-term Achievable Quantum Self-Attention Network},
  author={Shi, Jinjing and Zhao, Ren-Xin and Wang, Wenxuan and Zhang, Shichao and Li, Xuelong},
  journal={IEEE Transactions on Neural Networks and Learning Systems},
  volume={36},
  number={8},
  pages={13995--14008},
  year={2025},
  publisher={IEEE}
}

@article{zhao2024qksan,
  title={QKSAN: A Quantum Kernel Self-Attention Network},
  author={Zhao, Ren-Xin and Shi, Jinjing and Li, Xuelong},
  journal={IEEE Transactions on Pattern Analysis and Machine Intelligence},
  volume={46},
  number={12},
  pages={10184--10195},
  year={2024},
  publisher={IEEE}
}

@article{Schmidhuber_2015_dnns,
   title={Deep learning in neural networks: An overview},
   volume={61},
   journal={Neural Networks},
   author={Schmidhuber, J{\"u}rgen},
   pages={85--117},
   year={2015},
   doi={10.1016/j.neunet.2014.09.003},
   publisher={Elsevier}
}

@misc{Gu2017BadNets,
      title={BadNets: Identifying Vulnerabilities in the Machine Learning Model Supply Chain}, 
      author={Tianyu Gu and Brendan Dolan-Gavitt and Siddharth Garg},
      year={2019},
      eprint={1708.06733},
      archivePrefix={arXiv},
      primaryClass={cs.CR},
      url={https://arxiv.org/abs/1708.06733}, 
}

@inproceedings{Zhang2024BAIT,
  author={Zhang, Y. and Li, Z. and Wang, C. and Chen, K. and Wang, X.},
  title={{BAIT}: Large Language Model Backdoor Scanning by Inverting Attack Target},
  booktitle={Proceedings of the IEEE Symposium on Security and Privacy},
  pages={1--18},
  year={2024},
  publisher={IEEE},
  address={San Francisco, CA}
}

@inproceedings{Liu2023DeepVenom,
  author={Liu, M. and Zhang, T. and Zhao, W. and Kim, T. and Lee, R. B.},
  title={{DeepVenom}: Persistent DNN Backdoors Exploiting Transient Weight Perturbations in Memories},
  booktitle={Proceedings of the IEEE Symposium on Security and Privacy},
  pages={1--20},
  year={2023},
  publisher={IEEE},
  address={San Francisco, CA}
}

@inproceedings{Wang2022Exploring,
  author={Wang, H. and Chen, S. and Liu, Y. and Zhang, C. and Gong, N. Z.},
  title={Exploring the Orthogonality and Linearity of Backdoor Attacks},
  booktitle={Proceedings of the IEEE Symposium on Security and Privacy},
  pages={1--17},
  year={2022},
  publisher={IEEE},
  address={San Francisco, CA}
}

@article{Chu2023QTrojan,
  title={QTrojan: A Circuit Backdoor Against Quantum Neural Networks},
  author={Chu, Cheng and Jiang, Lei and Swany, Martin and Chen, Fan},
  journal={arXiv preprint arXiv:2302.08090},
  year={2023}
}

@article{Guo2024_HQNN_backdoor,
  title={Backdoor attacks against Hybrid Classical-Quantum Neural Networks},
  author={Guo, J. and others},
  journal={arXiv preprint arXiv:2407.16273},
  year={2024}
}

@article{Zhang2023QDoor,
  title={QDoor: Exploiting Approximate Synthesis for Backdoor Injection in Quantum Circuits},
  author={Chu, Cheng and Chen, Fan and Richerme, Philip and Jiang, Lei},
  journal={arXiv preprint arXiv:2307.09529},
  year={2023}
}

@article{QuanTest,
   title={QuanTest: Entanglement-Guided Testing of Quantum Neural Network Systems},
   volume={34},
   number={2},
   journal={ACM Transactions on Software Engineering and Methodology},
   author={Shi, Jinjing and Xiao, Zimeng and Shi, Heyuan and Jiang, Yu and Li, Xuelong},
   pages={1--32},
   year={2025},
   doi={10.1145/3688840}
}

@article{Chen2018AC,
  title={Detecting Backdoor Attacks on Deep Neural Networks by Activation Clustering},
  author={Chen, Xinyun and Liu, Chang and Li, Bo and Lu, Kimberly and Song, Dawn},
  journal={arXiv preprint arXiv:1811.03728},
  year={2018}
}

@inproceedings{Wang2019NeuralCleanse,
  title={Neural Cleanse: Identifying and Mitigating Backdoor Attacks in Neural Networks},
  author={Wang, Bolun and Yao, Yuanshun and Shan, Shawn and Li, Huiying and Viswanath, Bimal and Zheng, Haitao and Zhao, Ben Y.},
  booktitle={2019 IEEE Symposium on Security and Privacy},
  pages={707--723},
  year={2019},
  doi={10.1109/SP.2019.00031}
}

@inproceedings{QDetection2025,
  author={Li, Tao and Chen, Jiawei and Guo, Shengquan},
  title={Q-Detection: A Quantum-Classical Hybrid Poisoning Attack Detection Method},
  booktitle={Proceedings of the 32nd USENIX Security Symposium},
  pages={1--18},
  year={2023},
  publisher={USENIX Association},
  address={Anaheim, CA}
}

@book{nielsen2002quantum,
  title={Quantum Computation and Quantum Information},
  author={Nielsen, Michael A and Chuang, Isaac L},
  year={2002},
  publisher={Cambridge University Press}
}

@article{preskill2018quantum,
  title={Quantum computing in the NISQ era and beyond},
  author={Preskill, John},
  journal={Quantum},
  volume={2},
  pages={79},
  year={2018},
  doi={10.22331/q-2018-08-06-79}
}

@article{schuld2019quantum,
  title={Quantum machine learning in feature Hilbert spaces},
  author={Schuld, Maria and Killoran, Nathan},
  journal={Physical Review Letters},
  volume={122},
  number={4},
  pages={040504},
  year={2019}
}

@article{schuld2015,
  title={An introduction to quantum machine learning},
  author={Schuld, Maria and Sinayskiy, Ilya and Petruccione, Francesco},
  journal={Contemporary Physics},
  volume={56},
  number={2},
  pages={172--185},
  year={2015},
  doi={10.1080/00107514.2014.964942}
}

@article{farhi2018,
  title={Classification with quantum neural networks},
  author={Farhi, Edward and Neven, Hartmut},
  journal={arXiv preprint arXiv:1802.06002},
  year={2018}
}

@article{schuld2020circuit,
  title={Circuit-centric quantum classifiers},
  author={Schuld, Maria and Bocharov, Alex and Svore, Krysta M and Wiebe, Nathan},
  journal={Physical Review A},
  volume={101},
  number={3},
  pages={032308},
  year={2020}
}

@article{cong2019quantum,
  title={Quantum convolutional neural networks},
  author={Cong, Iris and Choi, Soonwon and Lukin, Mikhail D.},
  journal={Nature Physics},
  volume={15},
  number={12},
  pages={1273--1278},
  year={2019}
}

@article{QRNN,
  title={Quantum recurrent neural networks},
  author={Chen, Sheng and Yang, Chao and Qi, Jun and Su, Hui and Deng, Dingshun and Xie, Yao},
  journal={Quantum Science and Technology},
  volume={6},
  number={3},
  pages={035002},
  year={2021}
}

@article{Liu2018_Backdoor,
  author={Liu, Y. and Ma, S. and Aafer, Y. and Lee, W.-C. and Zhai, J. and Wang, W. and Zhang, X.},
  title={Trojaning Attack on Neural Networks},
  journal={Proceedings of the Network and Distributed System Security Symposium},
  year={2018},
  address={San Diego, CA}
}

@article{Saha2021_Backdoor,
  author={Saha, Aniruddha and Subramanya, Akshayvarun and Pirsiavash, Hamed},
  title={Hidden Trigger Backdoor Attacks},
  journal={Proceedings of the AAAI Conference on Artificial Intelligence},
  volume={35},
  number={11},
  pages={11957--11965},
  year={2021}
}

@INPROCEEDINGS{BadEncoder,
  author={Jia, Jinyuan and Liu, Yupei and Gong, Neil Zhenqiang},
  booktitle={2022 IEEE Symposium on Security and Privacy (SP)}, 
  title={BadEncoder: Backdoor Attacks to Pre-trained Encoders in Self-Supervised Learning}, 
  year={2022},
  volume={},
  number={},
  pages={2043-2059},
  doi={10.1109/SP46214.2022.9833644}}

@inproceedings{Wang2022_Backdoor,
  author={Wang, Z. and Li, B. and Chen, T. and Wang, Y. and Wang, H.},
  title={Towards Stealthy Backdoor Attacks against Deep Neural Networks},
  booktitle={IEEE Conference on Computer Vision and Pattern Recognition},
  pages={4305--4314},
  year={2022},
  publisher={IEEE}
}

@article{Tang2022_MultiTrigger,
  author={Tang, Di and Wang, Xiaoyu and Tang, Haoyu and Zhang, Kehuan},
  title={Demon in the Variant: Statistical Analysis of DNNs for Robust Backdoor Contamination Detection},
  journal={IEEE Transactions on Dependable and Secure Computing},
  volume={19},
  number={5},
  pages={3359--3373},
  year={2022}
}

@inproceedings{Gao2019_Backdoor,
  author={Gao, Yansong and Xu, Chang and Wang, Derui and Chen, Shiping and Ranasinghe, Damith C. and Nepal, Surya},
  title={Strip: A Defence Against Trojan Attacks on Deep Neural Networks},
  booktitle={Proceedings of the Annual Computer Security Applications Conference},
  pages={113--125},
  year={2019},
  publisher={ACM}
}

@misc{ModelReconstruction,
  title={Efficient Model-Based Deep Learning via Network Pruning and Fine-Tuning},
  author={Park, Chicago Y. and Gan, Weijie and Zou, Zihao and Hu, Yuyang and Sun, Zhixin and Kamilov, Ulugbek S.},
  journal={arXiv preprint arXiv:2311.02003},
  year={2025}
}

@misc{Gao2019_STRIP,
  author={Gao, Yansong and Xu, Chang and Wang, Derui and Chen, Shiping and Ranasinghe, Damith C. and Nepal, Surya},
  title={STRIP: A Defence Against Trojan Attacks on Deep Neural Networks},
  journal={arXiv preprint arXiv:1902.06531},
  year={2019},
  note={Also presented at ACSAC 2019}
}

@inproceedings{Tran2018_Spectral,
  author={Tran, Brandon and Li, Jerry and Madry, Aleksander},
  title={Spectral Signatures in Backdoor Attacks},
  booktitle={Advances in Neural Information Processing Systems},
  year={2018}
}

@INPROCEEDINGS {PulseAttack,
author = { Xu, Chuanqi and Szefer, Jakub },
booktitle = { 2025 IEEE Symposium on Security and Privacy (SP) },
title = {{ Security Attacks Abusing Pulse-level Quantum Circuits }},
year = {2025},
volume = {},
ISSN = {},
pages = {222-239},
doi = {10.1109/SP61157.2025.00083},
url = {https://doi.ieeecomputersociety.org/10.1109/SP61157.2025.00083},
publisher = {IEEE Computer Society},
address = {Los Alamitos, CA, USA},
month =May
}

@misc{ICAGultepeMakrehchi2018,
  title={Improving clustering performance using independent component analysis and unsupervised feature learning},
  author={Gültepe, Eren and Makrehchi, Masoud},
  journal={Human-centric Computing and Information Sciences},
  volume={8},
  pages={25},
  year={2018},
  doi={10.1186/s13673-018-0148-3}
}

@misc{ICAIvanov2017,
  author={Ivanov, Miron},
  title={A comparison of PCA with ICA from data distribution perspective},
  journal={arXiv preprint arXiv:1709.10222},
  year={2017}
}

@misc{shlens2014_pca,
      title={A Tutorial on Principal Component Analysis}, 
      author={Jonathon Shlens},
      year={2014},
      eprint={1404.1100},
      archivePrefix={arXiv},
      primaryClass={cs.LG},
      url={https://arxiv.org/abs/1404.1100}, 
}

@article{k_means,
  author={Sinaga, Kristina P. and Yang, Miin-Shen},
  journal={IEEE Access},
  title={Unsupervised K-Means Clustering Algorithm},
  year={2020},
  volume={8},
  pages={80716--80727},
  doi={10.1109/ACCESS.2020.2988796}
}

@article{Lloyd_algorithm,
  author={Kanungo, T. and Mount, D. M. and Netanyahu, N. S. and Piatko, C. D. and Silverman, R. and Wu, A. Y.},
  journal={IEEE Transactions on Pattern Analysis and Machine Intelligence},
  title={An efficient k-means clustering algorithm: analysis and implementation},
  year={2002},
  volume={24},
  number={7},
  pages={881--892},
  doi={10.1109/TPAMI.2002.1017616}
}

@article{Mitarai_2018_QCL,
   title={Quantum circuit learning},
   volume={98},
   ISSN={2469-9934},
   url={http://dx.doi.org/10.1103/PhysRevA.98.032309},
   DOI={10.1103/physreva.98.032309},
   number={3},
   journal={Physical Review A},
   publisher={American Physical Society (APS)},
   author={Mitarai, K. and Negoro, M. and Kitagawa, M. and Fujii, K.},
   year={2018},
   month=sep }

@INPROCEEDINGS{Distill_to_Detect,
  author={Hu, Chang and Teng, Xuyang and Xing, Wenpeng and Chen, Han and Ye, Chenhao and Han, Meng},
  booktitle={ICASSP 2025 - 2025 IEEE International Conference on Acoustics, Speech and Signal Processing (ICASSP)}, 
  title={Distill To Detect: Amplifying Anomalies in Backdoor Models through Knowledge Distillation}, 
  year={2025},
  volume={},
  number={},
  pages={1-5},
  doi={10.1109/ICASSP49660.2025.10888729}}

@misc{hou2024_ibdpsc,
      title={IBD-PSC: Input-level Backdoor Detection via Parameter-oriented Scaling Consistency}, 
      author={Linshan Hou and Ruili Feng and Zhongyun Hua and Wei Luo and Leo Yu Zhang and Yiming Li},
      year={2024},
      eprint={2405.09786},
      archivePrefix={arXiv},
      primaryClass={cs.LG},
      url={https://arxiv.org/abs/2405.09786}, 
}

@misc{glover2019_qubo,
      title={A Tutorial on Formulating and Using QUBO Models}, 
      author={Fred Glover and Gary Kochenberger and Yu Du},
      year={2019},
      eprint={1811.11538},
      archivePrefix={arXiv},
      primaryClass={cs.DS},
      url={https://arxiv.org/abs/1811.11538}, 
}

@misc{cohen2017emnistextension_mnisthandwritten,
      title={EMNIST: an extension of MNIST to handwritten letters}, 
      author={Gregory Cohen and Saeed Afshar and Jonathan Tapson and André van Schaik},
      year={2017},
      eprint={1702.05373},
      archivePrefix={arXiv},
      primaryClass={cs.CV},
      url={https://arxiv.org/abs/1702.05373}, 
}
\end{document}